%% file: 2Bells_2014n-arxiv.tex
\documentclass[12pt]{iopart}
\usepackage{graphicx,color,amssymb,amsfonts,hyperref}
\usepackage[frenchb,english]{babel}
\definecolor{nblue}{rgb}{0.3,0.3,1.0}
\definecolor{ngreen}{rgb}{0.2,0.7,0.2}
\definecolor{nred}{rgb}{0.9,0.1,0}
\definecolor{npurple}{rgb}{0.8,0.2,0.8}
\definecolor{golden}{rgb}{0.8,0.6,0.1}
\definecolor{nsilver}{rgb}{0.3,0.4,0.5}
\definecolor{nbrown}{rgb}{0.8,0.4,0.15}
\definecolor{nrose}{rgb}{0.7,0,0.35}
\definecolor{nviol}{rgb}{0.5,0,1.0}
\definecolor{nazur}{rgb}{0,0.35,0.7}
\definecolor{nchart}{rgb}{0.2,0.4,0}
\newcommand{\azur}{}
\newcommand{\blk}{\color{black}}
\newcommand{\grn}{}
\newcommand{\red}{}
\newcommand{\gold}{}

\newcommand{\beq}{\begin{equation}}
\newcommand{\eeq}{\end{equation}}
\newcommand{\bqa}{\begin{eqnarray}}
\newcommand{\eqa}{\end{eqnarray}}
\newcommand{\nn}{\nonumber}

\newcommand{\erf}[1]{Eq.~(\ref{#1})}

\newcommand{\ea}{{\em et al.}}

\newcommand{\ket}[1]{\left|{#1}\right\rangle}

\newcommand{\ie}{{\em i.e.}}
\newcommand{\eg}{{\em e.g.}}
\newcommand{\sq}[1]{\left[ {#1} \right]}
\newcommand{\cu}[1]{\left\{ {#1} \right\}}

\newcommand{\an}[1]{\left\langle{#1}\right\rangle}
\newcommand{\implies}{\Longrightarrow}

\newtheorem{theorem}{Theorem}

\newtheorem{axiom}[theorem]{Axiom}
\newtheorem{principle}[theorem]{Principle}
\newtheorem{assumption}[theorem]{Assumption}
\newtheorem{supposition}[theorem]{Supposition}

\newtheorem{corollary}[theorem]{Corollary}
\newtheorem{criterion}[theorem]{Criterion}
\newtheorem{definition}[theorem]{Definition}

\newtheorem{lemma}[theorem]{Lemma}

\newcommand{\EPR}{\mathcal{EPR}}
\newcommand{\Dis}{\mathcal{D}{\rm ist}}
\newcommand{\Rep}{\mathcal{R}{\rm ep}}
\newcommand{\Pre}{\mathcal{P}{\rm re}}

\newcommand{\Com}{\mathcal{C}{\rm om}}

\newcommand{\follows}{\Longleftarrow}
\newcommand{\et}{\wedge}

\begin{document}

\title[]{ The Two Bell's Theorems of John Bell}
\author{H. M. Wiseman}
\address{Centre for Quantum Computation and Communication Technology (Australian Research Council), Centre for Quantum Dynamics, Griffith University, Brisbane, Queensland 4111, Australia}

\begin{abstract}
Many of the heated arguments about the meaning of ``Bell's theorem" arise because this phrase can refer to two different theorems that John Bell proved, the first in 1964 and the second in 1976. 
His 1964 theorem is the incompatibility  of quantum phenomena with the dual assumptions  of locality and determinism. His 1976 theorem is the incompatibility  of quantum phenomena 
with the unitary property of local causality. 
This is contrary to Bell's own later assertions, that his 1964 theorem began with that single, 
and indivisible, assumption of local causality (even if not by that name). 
 While there are other forms of Bell's theorems --- which I present to explain the relation between 
Jarrett-completeness, ``\red fragile \blk locality'', and EPR-completeness --- I maintain that Bell's two versions are the essential ones. 
 Although the two Bell's theorems are logically equivalent, their assumptions are not, and the different versions of the theorem suggest quite different conclusions, which are embraced by different communities. 
 For realists, the notion of local causality, ruled out by Bell's 1976 theorem,  
is motivated implicitly by Reichenbach's Principle of common cause 
and explicitly by \red the Principle of relativistic causality, \blk and it is the latter which must be forgone.  
 Operationalists pay no heed to Reichenbach's Principle, but 
wish to keep \red the Principle of relativistic causality, \blk which, bolstered by an 
implicit  ``Principle of \red agent-causation\blk'',  implies their notion of locality. Thus 
for operationalists, Bell's theorem is the 1964 one, and implies  that it is determinism that must be forgone. 
I discuss why the two `camps' are drawn to these different conclusions,  
and what can be done to increase mutual understanding. 
\end{abstract} 

\maketitle

\section{Introduction} \label{sec:intro}

This special issue celebrates the 50th anniversary of both the submission and (surprisingly\footnote{Bell's paper  was received on 4 November 1964 and published in the November/December issue of that year \cite{Bel64}.   Presumably this issue was not {actually} printed until 1965, which presumption is supported by the fact that in many~\cite{Bel66,BelNau66,Bel76,Bel81,Bel86}
of his papers that referred to Ref.~\cite{Bel64}, Bell gave the year of its publication as 1965. Those with the date of 1964 may of course have been editorially ``corrected''.}) the publication of Bell's paper which introduced his eponymous theorem \cite{Bel64}. The present year is also the 50th anniversary of the submission of a less well-known paper by Bell~\cite{Bel66},  a prequel to the 1964 paper, which, due to editorial negligence, was not published until 1966~\cite{Mermin93}. This delayed prequel already contained a powerful impossibility theorem relating to hidden variables, a theorem which was formulated independently, and in an even stronger form, 
 by Kochen and Specker, in 1967~\cite{KS67}. 
  While the latters' names are now irrevocably associated with this contextuality theorem,  Mermin famously rescued Bell's earlier result from obscurity in his 1993 paper ``Hidden variables and the two theorems of John Bell" \cite{Mermin93}.

The present paper is also about two theorems proven by John Bell, but neither of them is that published in 1966. 
Rather, I will argue that, 
contrary to Bell's retrospective description of his own work~\cite{Bel81}, Bell proved two different 
forms of (what is known as) Bell's theorem. The first is that proven in 1964~\cite{Bel64},  
that there are quantum predictions incompatible with any theory satisfying locality and determinism (L\&D). 
The second is that proven in 1976~\cite{Bel76}, that there are quantum predictions incompatible 
with any theory satisfying local causality (LC). 
I use the terms {\em theorem} and {\em proven} advisedly. I am not concerned here with the intuitions 
that may have motivated Bell when he wrote his papers, but rather with what he rigorously proved 
(at least with rigour that should satisfy most theoretical physicists). 

Now while LC is a strictly weaker concept than L\&D (as will be discussed), 
it was highlighted by Fine in 1982~\cite{Fin82} that the range of phenomena  
respecting L\&D 
is the same as the range of phenomena respecting LC. Therefore Bell's 1976 theorem is a corollary of 
Bell's 1964 theorem,  as well as implying it.  Since the two Bell's theorems are thus logically 
equivalent, why should anyone but a pedantic historian of science distinguish them?  
The answer is: because the different theorems point in different directions. 
Bell's 1964 theorem suggests that Bell experiments leave us a with a choice:  
accept that physical phenomena violate determinism, or accept that they violate locality.  
Bell's 1976 theorem suggests that Bell experiments leave us with no choice: 
we must accept that physical phenomena violate local causality. 

These different interpretations of 
Bell's theorem are characteristic of different communities of physicists (which could be broadly characterised 
as operationalists and realists respectively). The failure to distinguish 
{\em which} Bell's theorem (1964 or 1976) one is discussing, and to 
use unambiguous terminology, is, I know from experience, a significant cause of miscommunication 
(see \eg~Ref.~\cite{blog13}). 
That is why I believe it is worthwhile for me to take the time to explain, and for the reader to take the time to 
understand, exactly what Bell proved, when he proved it, how he defined the terms he used to describe 
what he had proven, and how his terminology changed over time.  I also take the opportunity to examine 
alternate formulations of Bell's theorems, and to explore how neither L nor LC 
are immediate consequences of relativity theory, but rather require  at least one  additional assumption, 
different in the two cases.

Section~\ref{sec:notass} of this paper establishes the notation and assumptions that 
are common to most of the paper. In Section~\ref{sec:Bell64}  I examine the text of Bell's 1964 paper, 
give my reading of it in terms of locality and determinism, but also analyse alternate readings. 
In Section~\ref{sec:Bel76} I present Bell's 1976 paper, with its concept of local causality, 
and examine precedents for this concept. Section~\ref{sec:retrospec} looks critically at Bell's later  
representations of his own 1964 and 1976 theorems. Section~\ref{sec:related} covers 
related theorems by Fine and Jarrett, and introduces a new concept of ``\red fragile \blk locality''  
in order to relate Jarrett-completeness to EPR-completeness.  Section~\ref{sec:2camps} 
reviews the attitudes of the two ``camps" (operationalists and realists), how these
relate to Bell's two Bell's theorems, and  to the Principles of relativity, common cause, 
and ``\red agent-causation\blk''. Finally, in Sec.~\ref{sec:Disc}, I discuss why we cannot 
avoid considering at least Bell's two forms of Bell's theorem, 
and explain my recommendations to both communities regarding 
 how to cite, how to state, and how to construe, the appropriate Bell's theorem, 
if they wish for  productive discussions, or at least ``peaceful coexistence'' {\em \`a la} Ref.~\cite{Shi78}. 

\section{Notation and Assumptions} \label{sec:notass}

In this section I introduce the notation (largely following Bell~\cite{Bel64,Bel76,Bel71,Bel90b}) 
used throughout the paper and the assumptions which are implicit (or, in some papers, 
explicit) in Bell's theorems. 

\subsection{Notation} \label{sec:notation}

Consider two spatially separated and non-communicating observers, Alice and Bob, who choose to make particular measurements, and observe their outcomes.  Alice's choice of what measurement she is to make is represented by $a$. This is a free choice from some set $\cu{a}$.  For Bell's analysis~\cite{Bel64} of the EPR-Bohm scenario~\cite{Boh51}, $a$ was the direction along which to measure spin, and was notated $\vec{a}$. I am suppressing the vector notation 
for convenience and generality. 
The outcome of that measurement once she makes it is $A$.  The corresponding things for Bob are indicated by $b$ and $B$. For the most rigorous tests we should arrange that Alice's choice $a$ and outcome $A$ 
are space-like separated from Bob's $b$ and $B$. Here to avoid notational complexity 
I am using these symbols both for the values and for the events yielding them (\ie\ the 
event of Alice's making and implementing the choice $a$, and the event of the measurement's yielding the result $A$). 

In the common past light cone of $A$ and $B$ is $c$, an agreed-upon reproducible preparation procedure for the experiment which yields $A$ and $B$.  I denote by $\lambda$ any other variables (in addition to $c$, $a$, and $b$) that may affect the outcomes obtained. That is, they are variables not specified by the preparation procedure $c$, and as such may be deemed `hidden variables'. They may be specified in any region of space-time as long as it is earlier than the choices $a$ and $b$. Any space-like hypersurface may be chosen to define `earlier' here. 
(Note that there is no need to assume, as Bell did in some papers, 
that $\lambda$ corresponds to events in space-time. It could, for example, indicate a quantum pure state 
$\ket{\psi}$, which is a property of a hypersurface.)  

Using the above notation, we can define a phenomenon $\phi$ by the relative frequencies $f_\phi(A,B|a,b, c)$, 
where $c$ may be fixed, but $a$ and $b$ will, in the situations we are interested in, be allowed to vary. 
I use the term `relative frequencies' rather than `probabilities' to emphasise that a phenomenon 
is a feature of the world. However I will use the term `quantum phenomenon' for a phenomenon 
predicted by quantum mechanics, given how empirically succesful quantum mechanics is. Phenomena 
should be distinguished from {\em theories}. A theory $\theta$ for a phenomenon $\phi$  
comprises  at least  the following:  the set $\Lambda$ of values of $\lambda$; a mapping from $c$ to a probability measure $d\mu_\theta(\lambda|c)$ on $\Lambda$; and a specification of a joint probability distribution 
$P_\theta(A,B|a,b, c, \lambda)$, 
which predicts the phenomenon:  
\beq
\int_\Lambda d\mu_\theta(\lambda|c) P_\theta(A,B|a,b, c, \lambda) = f_\phi(A,B|a,b,c)
\eeq
Note that here, as everywhere in this paper, when an equation involving variables appears without quantifiers for those variables, a universal quantifier ($\forall$) is to be understood. 

 The set $\Lambda$ could contain a single $\lambda$, or the measure $d\mu_\theta(\lambda|c)$ could be singular, 
 or the dependences of $P_\theta(A,B|a,b, c, \lambda)$ on $\lambda$ could 
be trivial. In these cases the theory does nothing more than reproduce the phenomenon. Such theories can be called {\em operational} theories. A theory, operational or otherwise, might postulate other variables or events, 
in addition to $A$, $a$, $B$, and $b$, which are later than the space-like hypersurface introduced to define $\lambda$. 
 For the purpose of this paper, except for the discussion of Criterion~\ref{def:LC},  
 I will assume that this is not the case, and restrict consideration to those events listed. 

\subsection{Assumptions} \label{sec:Assumptions}
\begin{axiom}[Macroreality] \label{axiom:macro}
In an individual experimental run,  exactly one outcome $A$ and exactly one outcome $B$   
really happen, and are not `relative' to anything.  
\end{axiom}\vspace{-2ex}
\begin{axiom}[Space-time] \label{axiom:STC} 
The concepts of space-like separation, light-cones, space-like hypersurface 
{\em etc.}~can be applied unambiguously in ordinary laboratory situations. \end{axiom}
\begin{axiom}[Arrow of time] \label{axiom:arrow} 
 A cause can only be in  the   past  of its effect. 
\end{axiom} \vspace{-2ex}
\begin{axiom}[Free choice] \label{axiom:willy} $a$ and $b$ are freely chosen  (\ie~effectively uncaused),  and so are independent of $c$, of $\lambda$, and of each other. 
\end{axiom}
 While these axioms may be doubted (see Sec.~\ref{sec:2camps}), most discussions 
of Bell's theorem take them for granted and so will I, in this paper. 
There may be other implicit assumptions but they would be 
 at least as uncontroversial as the above. 

 Note that the term  `cause'  in Axiom~\ref{axiom:arrow} is here undefined; this axiom 
has force only when used in conjunction with other notions explicitly involving the idea of causation. 
In this context, I also specify the following:
\begin{principle}[\red relativistic causality\blk]\label{pr:relativity}
In relation to causation, `the past' is to be understood as `the past light cone'.
\end{principle}
\begin{principle}[manifest  \red agent-causation\blk]\label{pr:efficacy}
If a phenomenon exhibits an event that is statistically dependent on a freely chosen action, then that action is a cause of that event.
\end{principle} 
I call these `Principles' rather than an `Axioms' because I will entertain rejecting them, or at least 
closely related Principles. With regard to \red relativistic causality, \blk members of the `realist camp' identified in Sec~\ref{sec:intro} would, indeed, typically reject this Principle in response to Bell's theorem (see Sec.~\ref{sec:irrec}). 
Regarding  manifest  \red agent-causation\blk, while Principle~\ref{pr:efficacy} may seem indisputable,  at least one 
causal decision theorist~\cite{Joy07} rejects it (R.~Briggs, pers.~comm.).  
While this Principle  is not directly related to Bell's theorem 
(see Sec.~\ref{sec:irrec}), there is a stronger, and therefore 
more disputable, notion of \red agent-causation \blk (Principle~\ref{pr:ce}) which 
the `operationalist camp' seems impelled to embrace.

\section{Bell's 1964 Theorem} \label{sec:Bell64}

\subsection{ The ``almost universal'' reading} \label{sec:rep}

 My reading of Bell's 1964 paper~\cite{Bel64}, which is the reading, Bell later complained,  ``almost universally reported''~\cite{Bel81}, is that the theorem {\em proven} there is:  
\begin{theorem}[Bell 1964] \label{thm:Bell64}
There exist quantum phenomena for which there is no theory satisfying locality and determinism.
\end{theorem}
For theorems \red like this I will also \blk use the more succinct expression: 
quantum phenomena violate either locality or determinism. 

 Bell's meaning  for determinism in a theory is, I think, uncontroversial: 
\begin{definition}[determinism] 
A theory $\theta$ is deterministic, \ie~satisfies determinism ({\rm D}),  iff (if and only if) all probabilities $P_\theta(A,B|a,b, c, \lambda)$ are either zero or one. 
\end{definition}
That is, determinism means that the outcomes are given by functions $A_\theta(a,b,c,\lambda)$ and $B_\theta(a,b,c,\lambda)$.  In a more general setting, determinism means that an outcome 
$A$ is determined by the ``initial values''~\cite{Bel64} of the hidden variables (since they are specified 
prior to some space-like hypersurface) plus any subsequent, freely chosen, measurement settings 
(which are uncaused events by Axiom~\ref{axiom:willy}). 

Bell's meaning for ``locality'' in a theory is, by contrast, extremely controversial. 
Therefore  we must carefully examine how Bell defines it 
in this paper. 
In his Sec.~I (Introduction), a single paragraph which also serves as the abstract, Bell states (with my edits, 
introduced in the interests of brevity, indicated by curly brackets and ellipses)  
 \begin{quote}
\ldots\ Einstein, Podolsky and Rosen \ldots\ argu\{ed\} that quantum mechanics could not be a complete theory but should be supplemented by additional variables \{in order\} to restore to the theory causality and locality [2]. In this note that idea will be formulated mathematically and shown to be incompatible with the statistical predictions of quantum mechanics. 
\end{quote}
Here Bell identifies two concepts, causality and locality. He does not define causality at this (or any other) point, and the reference ``[2]" in the above quote is to a footnote which consists solely of 
a quote from Einstein,
\begin{quote}
But on one supposition we should, in my opinion, absolutely hold fast: the real factual situation of system $S_2$ is independent of what is done with the system $S_1$, which is spatially separated from the former.
\end{quote}
and its source, my Ref.~\cite{Ein49}. 

It is not immediately clear in the text whether Bell means this footnote to gloss ``causality and locality'', or just ``locality''. 
However, in the very next sentence Bell says  
\begin{quote}
\ldots\ the requirement of locality, or more precisely that the result of a measurement on one system be unaffected by operations on a distant system with which it has interacted in the past \ldots.
\end{quote}
That is, assuming that the ``real factual situation" of a system  is what is  probed by measuring it, 
Bell's definition of locality  follows from  the supposition of Einstein's which he 
quotes.  Now the notions of being ``independent of what is done with" or ``unaffected by 
 operations on" a system clearly refer to the action of an agent (say Alice) on her system, 
 and mean that Alice's action has no statistical effect. In the context of a Bell experiment, Alice's 
 action on her system is to perform some particular measurement $a$, chosen by her. 
 Thus we can formalise Bell's definition of locality, provisionally 
 (in the sense that we will have to test this definition against
what Bell says elsewhere in his paper), as follows: 
\begin{definition}[locality] \label{def:local}
A theory $\theta$ is local, \ie~satisfies locality ({\rm L}), iff 
\beq \label{local}
P_\theta(B|a, b, c, \lambda) = P_\theta(B|b, c, \lambda),
\eeq
plus the corresponding equation $P_\theta(A|a, b, c, \lambda) = P_\theta(A|a, c, \lambda)$ 
for Alice. 
\end{definition}
The ``corresponding equation for Alice" will go unstated in definitions from now on. Note that it is implicit in 
the above equation that the function $P_\theta(B|b, c, \lambda)$, which we have 
not previously considered, exists. 

As yet, what Bell means by causality remains unexplained. However, Bell closes Sec.~I with these statements: 
\begin{quote}  
 It is the requirement of locality \ldots\ that creates the essential difficulty. \ldots\  There have been  attempts \ldots\ to show that even without such a separability or locality requirement
no ``hidden variable'' interpretation of quantum mechanics is possible. \ldots\ \{However,\}   a hidden variable interpretation of elementary quantum theory [5] has been explicitly constructed. That particular interpretation has indeed a grossly non-local structure. This is characteristic, according to the result to be proved here, of any such theory which reproduces exactly the quantum mechanical predictions.
\end{quote}  
This passage demonstrates a number of points with Bell's terminology. First, he 
uses ``separability" and ``locality" interchangeably (as he had done also in his earlier paper \cite{Bel66}), 
though he uses the latter more often so it is the term I have adopted.  
Also, rather than the two assumptions of causality and locality, he states the two assumptions of 
``hidden variables'' and locality. In Ref.~\cite{Bel66}, Bell used ``hidden variables'' to mean  
\guillemotleft ``dispersion free'' states\guillemotright\ which make individual measurement results 
``determined'', and I am sure that is what he intends here also.  Now Bell clearly implies that 
for Einstein's program, locality is the problem, not causality, because of ``[5]'', which is 
Bohm's pilot wave interpretation~\cite{Boh52}, a perfectly fine (albeit nonlocal) 
deterministic hidden variable theory.  
All of this suggests to me that by ``causality'' Bell means nothing more 
than determinism, and was, again, simply being promiscuous with his terms. That is, Bell's {\em theorem}, 
as he describes in his final sentence, is that any deterministic theory of quantum phenomena must be nonlocal.  

 Bell uses the terms ``causality'' only a single time in the paper. Elsewhere he is clear that the second assumption 
(\ie\ in addition to locality) he requires for his theorem is determinism. In the last sentence of Bell's Sec.~V, he states: ``the statistical predictions of quantum mechanics are incompatible with separable predetermination."   (In the spirit of Axiom~\ref{axiom:arrow}, Bell makes no distinction between {\em pre}determined outcomes and determined outcomes.) 
The first sentence of Bell's Sec.~VI (Conclusion) reiterates this  statement of his theorem,  but with 
precise descriptions  rather than with any of the above terms:  
\begin{quote}
In a theory in which {\em parameters} are added to quantum mechanics to {\em determine the results of individual measurements}, without changing the statistical predictions, there must be a mechanism whereby {\em the setting of one measuring device  can influence the reading of another instrument, however remote}.
\end{quote}
Here the italics are my addition, to highlight Bell's definitions of determinism 
and (the negation of) of locality.

 As the above quote shows, Bell definitely means locality specifically as the absence of any influence of the {\em setting} $a$ of the remote measurement  device. This confirms my above reading of his definition 
of locality in Definition~\ref{def:local}. In fact this reading is confirmed in two more places in the paper. 
 In the first paragraph of Sec.~II, he states it in the context of an experiment on a spin-singlet state of two spin-half particles: 
\begin{quote}
we make the hypothesis [2] \ldots\  
that if the two measurements are made at places remote from one another, the orientation of one magnet does not influence the result obtained with the other.
\end{quote}
Recall that Bell's note [2] is the quote from Einstein, which Bell reuses here, and which, as 
I have shown above, Bell  associates  with locality. Later in that same section, he again states 
\begin{quote}
The vital assumption [2] is that the result 
$B$ for particle 2 does not depend on the setting $a$ of the magnet for particle 1, nor $A$ on $b$. 
\end{quote}
Finally, in (my) Ref.~\cite{Bel66}, Bell's prequel to his 1964 theorem,  his use of  
locality is consistent with the above,  as he elucidates 
the ``grossly nonlocal character" of Bohm's theory  \cite{Boh52} by showing that ``in this theory 
\ldots\ the disposition of one piece of apparatus affects the results obtained with a 
distant piece."

With this understanding of Bell's assumptions, the structure of his paper is as follows. 
His Section II is titled ``Formulation'', and this is where he gives the 
{\em mathematical formulation} (as promised in his Introduction) of the dual assumptions 
of causality (i.e. determinism) and locality,  his equation (1),
$$A(a,\lambda) = \pm 1, B(b,\lambda) = \pm 1.$$
In the context (spin measurements), the existence of these functions and their possible values, 
does indeed capture these dual 
assumptions\footnote{In this paper, 
Bell does not use $c$, which is why it is absent as an argument in these functions. 
However $\lambda$ can always be assumed to include $c$ for the 
purpose of characterising a theory as local or deterministic.}. 

Now prior to this formulation, Bell gives a one-paragraph motivation for considering hidden variable 
theories, by reiterating the EPR argument in its spin-singlet version~\cite{BohAha57}. The crucial 
sentence, which follows Bell's statement of the assumption ``the orientation of one magnet does not influence the result obtained with the other'' (quoted above), is 
\begin{quote}
Since we can predict in advance the result of measuring any chosen component of $\vec{\sigma}_2$  by previously measuring the same component of $\vec{\sigma}_1$, it follows that the result of any such measurement must actually be predetermined.
\end{quote}
 Here Bell has made a mistake. His conclusion (predetermined results) does not follow from \red his premises (predictability, and 
 the hypothesis stated in the preceding sentence). \blk This is simple to see from the following counter-example. Orthodox 
quantum mechanics (OQM) is a theory in which the setting $a$ of one device does not statistically influence the result $B$ obtained with the other:  
\beq 
P_\theta(B|a,b,c,\lambda) = P_\theta(B|b,c,\lambda).
\eeq
Here, if $c$ were to correspond to preparation of a mixed quantum state $\rho_c$,   
the variable $\lambda$ would allow for a pure-state decomposition; in purely operational QM, 
only $c$ would appear\footnote{In the present case, where $\rho_c$ is a singlet state (which is pure), there is no distinction.}.  But in OQM it is of course not true that the results of spin measurements are predetermined for a singlet state as Bell is considering. 

 Now, as stated, Bell's first paragraph (not counting the abstract) 
serves only to motivate the formulation of the theorem. This is to be expected from 
its placement prior to the mathematical formulation. It is also reflected in how the paragraph ends: 
``this predetermination implies the possibility of a more complete specification of the state." 
Having thus introduced the {\em possibility} of hidden variable (\ie~deterministic) 
theories, he then formulates such theories, in the next paragraph, 
under the hypothesis of locality, as per his equation (1): 
\begin{quote}
Let this more complete specification be effected by means of parameters $\lambda$. \ldots\ The result $A$ of measuring $\vec\sigma\cdot a$ is then determined by $a$ and $\lambda$, 
and \{similarly\} the result $B$ \ldots, and \{his equation (1)\}.  The vital assumption [2] is that the result 
$B$ for particle 2 does not depend on the setting $a$ of the magnet for particle 1, nor $A$ on $b$. 
\end{quote}
The end of this quotation is the locality assumption already quoted above.  Finally, that Bell's 
EPR paragraph forms no part of his 1964 theorem (which in this paper he calls his ``result'') 
is clear from the fact that after he has {\em formulated} the problem 
in Sec.~II, and {\em illustrated} it in Sec.~III, he finally gets to Sec.~IV where 
``The main result will now be proved.'' 

Thus I would classify Bell's mistake in this paragraph as 
a peccadillo, having no impact on the main result in his paper. 
 It would have been an easy mistake for Bell to have made, if he had the idea 
that EPR had already proven determinism from some sort of locality
assumption, and did not think hard about whether it was the same 
as the locality assumption he was about to use in his own theorem.  
Indeed the paper could be made completely sound simply by replacing ``it follows"
in the above (``Since we can predict \ldots'') quote by  ``the obvious explanation is'', or 
``EPR's premises imply".   Although Bell believed that 
he was reproducing EPR's argument,  EPR's premises (which are
never stated by Bell)  are {\em not} equivalent to locality 
 (as defined here by Bell), and they {\em do}  justify  
 the conclusion of predetermined outcomes; see Ref.~\cite{Wis13} and \ref{App:EPR:Com}.

\subsection{Alternate Readings} \label{sec:pec}

As the beginning of the preceding subsection implies, in later life Bell represented his 1964 paper 
differently from how it was ``almost universally reported'', as I will discuss in Sec.~\ref{sec:retrospec}. 
There is also a community of Bell's followers~\cite{DGZ92,Mau94,Nor06} who support his 
stance; see Sec.~\ref{sec:realists}. Their reading of Bell's paper differs from the usual understanding as follows. 
First, they see the first paragraph of Bell's ``Formulation'' 
section as an essential part of his 1964 theorem, the first part of a two-part argument. Second, they see 
this first part as legitimately deriving determinism from predictability and the assumption of locality, 
since they see Bell's concept of locality as being more general than my reading of it. 
Basically, they see Bell's notion of locality as being local causality (see next section), 
\red which requires an outcome to be statistically \blk 
independent not only of the remote setting, but also of the remote outcome. 
In particular, they do not see operational quantum mechanics as a counter-example whereby predictability 
and locality do not imply determinism. Thus they represent Bell's 1964 theorem as being that quantum phenomena 
violate the assumption of locality.  

This reading of course has the advantage that it does not require the reader to assume that Bell made a mistake, 
however inconsequential,  in his epitome of the EPR argument. 
It also has another advantage (Maudlin, pers.~comm.): Bell's second sentence,  
which talks of Einstein's hope that hidden variables would ``restore to the theory causality and locality [2]'' 
also implies that Bell believed that OQM did not satisfy the locality condition, or at least that Bell 
believed that Einstein believed this. 
If this is meant to reflect Bell's beliefs then it is, on my reading, the same mistake he made in reproducing the 
EPR argument, and is equally unimportant here. Regarding Einstein's beliefs, I will return to these  
in Sec.~\ref{sec:eb}, since Bell's definition of locality does not exactly coincide with 
the quote from Einstein he uses here. 

To me, the advantages of this reading are demonstrably outweighed by its many disadvantages: 
i) it does not explain why Bell would, in 1964,  state his result {\em four times} as requiring two assumptions, 
locality and determinism, and not once as requiring only the assumption of locality; 
ii) it does not explain why in his first subsequent paper on the topic of hidden variables~\cite{Bel71}, 
after seven years to think about how best to explain his result, he still states it 
(somewhat redundantly) as being ``that no local deterministic hidden-variable theory 
can reproduce all the experimental predictions of quantum mechanics''~\cite{Bel71}; 
iii) it does not explain why Bell would, in 1964, define locality {\em four times} in terms of independence 
from the remote setting, as per Definition~\ref{def:local}, and 
never any other way\footnote{In particular, there is no evidence to support the suggestion (Norsen, pers.~comm.) that 
Bell began with a general notion of locality, along the lines of local causality, and only 
narrowed to this definition after he had established determinism via his EPR paragraph. 
Bell's statement of the locality hypothesis in the middle of that paragraph is exactly 
the same as elsewhere in the paper.}; 
iv) it does not explain why Bell would state the conclusion of the supposedly crucial first part of his theorem as
being merely that it ``implies the possibility of a more complete specification of the state.''; 
v) it does not explain why Bell would place this supposedly crucial first part {\em prior} to the mathematical formulation 
of his result, and not mention it anywhere else in the paper. 

In short, to claim that the theorem Bell proved in his 1964 paper rules out  
a unitary `localistic' hypothesis (to coin a word),  
despite the fact that he never articulates a hypothesis that would \red be adequate to the task, \blk  
and invariably states his result as also requiring the assumption of determinism,  
is to do violence to the notions of theorem and proof. 
One might be tempted to venture that Bell never intended to provide a rigorous argument from precisely stated
premises to the sought-for contradiction with quantum mechanics, 
and so any one is  free to take his heuristic 
argument and make it rigorous in whatever way seems best. 
This is unfair to Bell, as he certainly 
does give such a rigorous argument, from the premises of D and L. Furthermore, Bell clearly did intend  
his argument to be rigorous: 
having previously \guillemotleft found wanting \{others'\}  {\em attempts} to show that \ldots\ no ``hidden variable" interpretation of
quantum mechanics is possible\guillemotright , he contrasts these attempts with ``the result to be {\em proved} here" 
(my emphasis).  

Of course  one is  free to  construct a proof using the EPR paragraph of Bell's 
paper to derive L and D,  if one corrects it by beginning with a sufficiently strong localistic 
hypothesis H.   This may be useful to obtain a pedagogical proof of Bell's theorem (in a general sense), as 
in Ref.~\cite{Nor11} for example, but that should not be confused with  Bell's 1964 theorem, 
or its proof.  In this context, one could ask what assumption H Bell may, hypothetically, have had in mind,  
which prompted him to give the motivating EPR argument, even if he failed to articulate it, for whatever
reason.  Readers not interested in this exploration may skip to Sec.~\ref{sec:Bel76}.

\subsection{Hypothetical hypotheses} \label{sec:eb}

In this subsection, I try to identify a localistic 
hypothesis H which is unambiguous, is sufficiently powerful to make the EPR argument work, is not 
tantamount to assuming D and L,  and plausibly could have been held by Bell in 1964. 
As we will see, even with this strictly limited aim, there are difficulties (in particular due to 
the final clause) with all three candidates that present themselves.  

The first obvious candidate for H is local causality (LC), 
a notion which Bell later adopted, as will be discussed in the following section.  Although  the notion of LC had been formulated mathematically 
 by the logician Boole as early as 1854 
(see Ref.~\cite{Pit89}), and by the philosopher Reichenbach in 1956~\cite{Rei56}, there is no reason to suspect 
that it was known to Bell in 1964\footnote{Intriguingly \cite{CavLal13}, Reichenbach 
 ---   whose  ``principle of common cause" for correlations~\cite{Rei56} is, given the axioms of 
Sec.~\ref{sec:Assumptions}, equivalent to LC~\cite{vFr82} --- had previously 
considered (and dismissed) hidden variable theories in 1944. That he did not combine these areas of knowledge
to prove Bell's 1976 theorem indeed ``goes to show how subtle and deep was Bell's insight" \cite{CavLal13}.}.  
In physics, the concept was developed incrementally, by Bell and others, in the first half of 
the 1970s (see Sec.~\ref{sec:prec}). 
 The two papers which Bell references in his EPR paragraph, 
Einstein's  autobiographical notes \cite{Ein49}, and the paper on the ``Paradox of Einstein, 
Rosen, and Podolsky'' ({\em sic.}) by Bohm and Aharanov~\cite{BohAha57} both use the  same sort of  
localistic notion Bell does,  referring only to the influence of the remote setting.  
 The quote from Ref.~\cite{Ein49} 
has already been examined\footnote{It must be admitted, however, that if 
Bell had chosen a {\em different} quote from Ref.~\cite{Ein49}, {\em viz.} ``the real situation of $S_2$ must be independent of what {\em happens to} $S_1$'' --- my emphasis --- then my case would not have been so clear cut; 
see Sec.~\ref{sec:realists} for the nuances of word choice in this context. One should not read too much 
into Einstein's diction in the quote in this footnote, however. Writing about the same time on ``quantum mechanics and reality'' \cite{Ein48}, Einstein elevates the ``principle of contiguity'', saying that without it empirical physics would be impossible. Einstein's definition of this principle is practically the same as the other quote which Bell used to motivate his definition of  locality: ``\{For\} objects far apart in space (A and B): external influence on A has no direct influence on B.''}  and will be further discussed below.  The assumptions of Ref.~\cite{BohAha57} are wholly implicit, 
but the only notion of locality there is in the negative: a hypothetical nonlocal  ``hidden interaction'' 
which ``would  have to be instantaneous, because the orientation of the measuring apparatus \{for $A$\} could very quickly be changed, and the spin of $B$ would have to respond immediately to the change.''  

 The second candidate for H is   
 {\em completeness}, as EPR~\cite{EPR35} defined it, or, rather, characterised it, using a web of several 
conditions, necessary or sufficient as required; see Ref.~\cite{Wis13} and \ref{App:EPR:Bel} of the current paper. 
As shown there, the notion of completeness is sufficient to derive a contradiction with 
the predictions of quantum mechanics, using the arguments in Bell's 1964 paper. 
 The EPR paper was unquestionably  
known to Bell, and he even entitled his paper ``On the Einstein Podolsky Rosen Paradox''. 
However, as just indicated, the EPR notion of completeness is far from simple in logical terms, and 
there is no evidence that Bell had  a formal understanding of it.  
He makes no mention of any of the formal criteria EPR introduce,  
and  does not use ``complete" in their technical sense,  talking, for example, of a ``more complete 
specification of a state."   He never quotes from the EPR paper~\cite{EPR35}, preferring Einstein's  
 less technical notes \cite{Ein49}, and in his prequel~\cite{Bel66},  
the latter was his {\em only} reference for the ``Einstein-Podolsky-Rosen paradox.''   
Bell's other inspiration in 1964, the paper by Bohm and Aharanov~\cite{BohAha57}, 
is even less formal than Ref.~\cite{Ein49}. Despite its title, it is disconcertingly disengaged from the EPR paper. 
It does not avail itself of EPR's terminology, mentioning neither ``completeness'' nor ``reality'',  
and ubiquitously gives the three authors' names in the wrong order.    

 The final candidate for H is the supposition Bell quotes from Ref.~\cite{Ein49}, already mentioned above. 
It is worth requoting it here:
\begin{supposition}[No telepathy] \label{sup:NT}
The real factual situation of system $S_2$ is independent of what is done with the system $S_1$, which is spatially separated from the former.
\end{supposition}
(For the name I have given this supposition, see the quote below.)
Now although Bell seemed to indicate (twice) that this was equivalent to his definition of locality, it is different  
in that it requires not that Bob's {\em result} $B$ be independent of Alice's setting $a$, but rather that the  
``real factual situation'' of Bob's system be thus independent\footnote{In this, Supposition~\ref{sup:NT} arguably has a similar 
relation to locality as per Definition~\ref{def:local} as does local causality to factorizability; see Sec.~\ref{sec:prec}. 
For a couple of reasons, which this footnote is too small to contain, I suggest ``separability'' as an 
appropriate name for the exact analogue to local causality.}. 
Einstein never states what qualifies something to be a ``real factual situation''. 
With some thought, one can propose a strong but reasonable formalisation for this concept 
such that the assumption (which, prior to the passage quoted \red above, \blk Einstein makes explicitly) that systems have 
real factual situations, plus the no telepathy supposition, has the same 
force as local causality (see Criterion~\ref{def:LC}). However, there is no suggestion of such a formalisation 
in Ref.~\cite{Ein49}. Nor does Einstein use it there to make the argument that Bell wants to make, 
from predictability to determinism. 
Even if Bell had worked out such a formulation from studying Einstein's text,  
Supposition~\ref{sup:NT} alone would not be strong enough to do its 
hypothetical role in Bell's EPR paragraph. As Bell would have known if he had made such a study, 
Einstein readily admitted that the prior assumption mentioned above is necessary: 
\begin{quote}
One can escape from this conclusion [that statistical quantum theory is incomplete] only by either assuming that the measurement of $S_1$ (telepathically) changes the real situation of $S_2$ or by denying independent real situations as such to things which are spatially separated from each other. Both alternatives appear to me equally unacceptable.
\end{quote}
 
Leaving aside any hypothetical formulation of of Einstein's 
``real factual situation'' that could have worked for Bell, it is interesting 
to ask the question whether Supposition~\ref{sup:NT} alone 
is violated by OQM (as Bell implies). 
 Einstein certainly  argues in Ref.~\cite{Ein49} that this is so,  
if, as ``it appears to me'', ``one may speak of the real factual situation 
 of the partial system $S_2$'' even when one takes OQM to be complete. That is, in Einstein's understanding, 
\red if quantum mechanics is complete then \blk 
  the quantum state of $S_2$,  conditioned on measurements on $S_1$, must be a 
``real factual situation''. A   `realist' understanding  of OQM does involve such a physical collapse of 
the wavefunction, which requires specifying some foliation of space-time and does indeed violate Einstein's supposition. 
 However strictly operational QM does not violate Supposition~\ref{sup:NT}, because 
strict operationalists do indeed operate ``by denying independent real situations as such'', as Einstein subsequently allowed. 
Thus even if Bell did have ``no telepathy'' in mind for locality, 
his implication that OQM violates locality would be incorrect, because OQM can be interpreted purely operationally. 
It is also important to remember that OQM satisfies locality 
as Bell actually expresses it, as per Definition~\ref{def:local}, \red regardless of whether 
one has a realist or an operationalist understanding of the theory. \blk

\section{Bell's 1976 theorem} \label{sec:Bel76}

 As mentioned above, in 1971 Bell wrote another (\ie~in addition to Ref.~\cite{Bel66}) review of the ``hidden-variable question"  \cite{Bel71}, this time with the opportunity to give his own result its due emphasis.  
With characteristic modesty, he co-referenced his own 1964 paper with Refs.~\cite{CHSH69,Wig70}, 
which closely followed his own presentation, relying upon (deterministic) 
``hidden variables" with the ``locality" condition or assumption. 
As will be discussed below, the presentation in Ref.~\cite{Bel71} does show  
some evolution in Bell's thinking, but a clear conceptual advance is not to be found until his 1976 paper~\cite{Bel76}, 
where he  introduced  the term ``local causality" (LC).

Reference~\cite{Bel76} made use of the concept ``local beable", 
Bell's term (introduced in Ref.~\cite{Bel73}) 
for elements of reality that were assumed to be localisable to points, or at least small regions, 
in space-time. Many potential elements of reality are not usually conceived this way 
--- a quantum state, for example, does not exist in space at all, but rather is defined on a space-like hypersurface~\cite{Mau94}. As Bell himself said in 1981 \cite{Bel81}, ``These variables \{($\lambda$)\} could well include, for example, quantum mechanical state vectors, which have no particular localization in ordinary space time." 
However, one could, if one wished, imagine that every point on that space-like hypersurface carries a representation of the quantum state, and in that way one could construct an ontology of the sort Bell wished for in 1976. 
 Thus we can continue to use   Bell's earlier notation of Ref.~\cite{Bel64}, to which 
he returned in his definitive presentation on the subject \cite{Bel90b}, just before he died. 
 The conceptual advance which the notion of local beables prompted for Bell was to treat 
settings and outcomes (and other, postulated, beables) on the same footing.  

 Thankfully, unlike the case for Ref.~\cite{Bel64}, it is not necessary to do a detailed textual analysis to convince 
 anyone of the premise Bell introduced in 1976\footnote{That said, one may become confused if one tries to follow his argument  from his equations and figures alone, since he introduces in his Eq.~(2) the notion of local causality as the statistical irrelevance of beable $B$, but applies it to his Eq.~(5) as the statistical irrelevance of beables $M$ and $B$.}. He first mentions the idea that theories of  classical physics  exhibit {\em local determinism}. 
  Being formulated in terms of beables, this is more general than his considerations in 1964, 
 but it certainly implies the joint premise of L\&D he made there.  Then he says 
 \begin{quote}
 We would like to form some notion of {\em local causality}, in theories which are not deterministic, 
 in which the correlations prescribed by the theory, for the beables, are weaker. 
 \end{quote}
 Using the usual notation, Bell's new concept can be characterised as follows: 
\begin{criterion}[local causality] \label{def:LC}
A theory $\theta$ is locally causal, \ie~satisfies local causality ({\rm LC}),  only if  
\beq \label{LC}
P_\theta(B|A, a, b, c, \lambda) = P_\theta(B|b, c, \lambda).
\eeq
\end{criterion} \grn
Note that this is a necessary criterion (`only if', not `iff') because here Bell's general notion 
is applied only to the specific beables relating to  hidden variables ($\lambda$) prior 
to some particular space-like hypersurface, and settings and outcomes posterior to it.  

The idea  of local causality  is that any events, such as $A$ and $a$, which lie outside 
the past light cone of event $B$,  are  statistically irrelevant 
to $B$. \red Of course, one would only expect LC to hold if one's theory  
included a description of all prior variables $\lambda$ relevant to the outcomes; 
otherwise outcome $A$ could contain information about $\lambda$ which 
would make it statistically relevant to outcome $B$. 
Note, however,  that the presence of $\lambda$ does not imply \blk any assumption of 
determinism here. That is, the concept of LC may apply even for non-deterministic 
theories such as OQM. Indeed, when $\rho_c$ is  an unentangled state, 
there is always an instance of OQM (corresponding to a decomposition of $\rho_c$ 
into pure product states~\cite{Nielsen00a}) which satisfies LC. 
But for entangled states,  orthodox {\em quantum mechanics is not locally causal}, 
as Bell says as his 3rd section heading of Ref.~\cite{Bel76}. 
Moreover, using assumption (\ref{LC}), one can (and Bell does) derive an 
inequality which contradicts quantum predictions, thus proving 
\begin{theorem}[Bell 1976] \label{thm:Bell76}
There exist quantum phenomena for which there is no theory satisfying local causality.
\end{theorem}
To put it another way,  ``quantum mechanics is {\em not} embeddable in a locally causal theory" 
\cite{Bel76}. 

  \setcounter{footnote}{0}

It is unfortunate (for me writing this article, and for anyone reading  
Ref.~\cite{Bel76} subsequent to his more famous 1964 paper) that, almost as soon as he had introduced the term LC, 
Bell started using the term ``locality" for the same thing, referring to a ``locality inequality" 
(for what everyone else calls a Bell inequality) and the ``locality hypothesis" in Ref.~\cite{Bel76}.   
 However, at least in his last word on the subject \cite{Bel90b}, Bell showed his preference 
 unequivocally.  Here Bell  
 carefully discusses various answers to the question ``what cannot go faster than light?", 
 and when referring to the concept
 \ref{def:LC}, he uses the term ``local causality" at least 17 times, and the term ``locality" only twice 
 (in a single paragraph discussing experiments which is rather disconnected from the rest of the paper)\footnote{He does however use, quite deliberately, the term ``locally explicable" as synonym for ``locally causal" with reference to correlations.  It would be a mistake to regard the ``explicable" here as inessential. The concept of LC is all about explaining correlations, as Bell goes to great 
 lengths to explain in Ref.~\cite{Bel90b} and elsewhere.  The idea of ``explaining'' correlations is formalised  in Principle~\ref{pr:cc}.}.  Thus I believe it respects Bell's final will (and avoids confusion when reading his pre-1976 papers) to 
 use only the term LC for this concept, and to ignore Bell's occasional lapses in this regard.

 \subsection{Precedents for Bell's concept of Local Causality} \label{sec:prec}
 
 While Bell certainly introduced, and promoted, the term LC for the concept \ref{def:LC}, 
 there is a brief history, in the context of quantum phenomenology, of similarly  formulated  mathematical 
conditions. 
In particular,  in 1974 Clauser and Horne introduced the  
 notion  of a ``factored form''~\cite{CH74} 
 of the joint probability distribution:
 \begin{definition}[factorisability]  \label{def:factor}
 A theory $\theta$ is factorisable, \ie~satisfies factorisabilty ({\rm F}), iff 
 \beq \label{factor}
   P_\theta(A,B|a,b,c,\lambda) =    P_\theta(A |a, c,\lambda)   P_\theta( B| b,c,\lambda).
 \eeq
 \end{definition}
 Equation~(\ref{factor}) is mathematically equivalent to \erf{LC} (with the corresponding equation for Alice of course).  
Clauser and Horne describe it as ``a reasonable locality condition" and 
 an ``extrapolation of the common-sense view that there is no action at \{a\} distance." They go on 
 ``\{W\}e call any theory in which it holds an objective local theory.$^{15}$" 
 A principal result of their paper is that 
 ``the predictions of objective local theories and of quantum mechanics differ."
 
 On the basis of the above, one could argue that Theorem~\ref{thm:Bell76}, which I have called 
 the 1976 Bell theorem, should be called the 1974 Clauser-Horne theorem. However, 
  Bell considered his formulation to be more fundamental, saying \cite{Bel90b}
  \begin{quote}
  Very often \ldots\ factorization is taken as the starting point for the analysis. Here we have preferred to see it not as the {\em formulation} of ``local causality", but as a consequence thereof.
  \end{quote} 
  In the sense that Bell defined local causality generally, in terms of beables rather than outcomes 
  and settings {\em etc.}, Bell's point is an important one. 
 Bell's terminology is also to be preferred, as ``locally causal" has a  unique  associated noun (local causality), 
 whereas ``objective local", comprising two adjectives, does not,  and instead  
 suggests two separate assumptions, 
 objectivity and locality. 
 The conceptual advance of Bell's 1976 theorem lies in 
the unitarity of the notion of local causality, so I believe it is appropriate to give the credit to Bell. 

As a final point,  
 Clauser and Horne themselves say, in their  footnote ($^{15}$) to the above quote, 
\guillemotleft The class of OLT \ldots\ is essentially the class of stochastic hidden-variable theories ``with a certain local character" considered by Bell (Ref. 4).\guillemotright\   
 Here ``Ref.~4'' refers to Bell's 1971 paper~\cite{Bel71}, so the credit for the theorem of Clauser and Horne 
 arguably lies with Bell in any case. 
Indeed, Bell's 1971 paper gives the general expression for a correlation function allowing (in a footnote) 
for ``indeterminism with a certain local character". Using the notation of Sec.~\ref{sec:notation}, Bell's expression is  
\beq \label{certainlocal}
\an{AB}_{a,b,c} = \int_\Lambda d\mu_\theta(\lambda|c) \bar{A}_\theta(a,\lambda,c) \bar{B}_\theta(b,\lambda,c),
\eeq
where $\bar{A}_\theta$ and $\bar{B}_\theta$ some real-valued functions specified by the theory, 
which give the {\em average} over local indeterminism not described by $\lambda$. 
Since $a$ or $b$ can correspond to arbitrary observables, 
including the identity, the above characterisation does capture all possible statistics of the measurement results $A$ and $B$ in a theory ``with a certain local character".  This phrase is, unfortunately, not fit to appear in a theorem, and Bell offers nothing better in Ref.~\cite{Bel71}. Thus, again, it seems better to stick with Bell's far more elegant notion of LC, and to keep my nomenclature for Theorem~\ref{thm:Bell76}.

\section{Later Commentary by Bell} \label{sec:retrospec}

 As noted in Sec.~\ref{sec:pec}, with the benefit of hindsight, 
Bell represented the logic of his 1964 paper~\cite{Bel64}  
rather differently from how it appears to me. He had, as just mentioned in 
Sec.~\ref{sec:prec}, considered  
``indeterminism with a certain local character" in 1971, 
and in a commentary from 1972 \cite{Bel72} he wrote: 
\begin{quote}
It can \ldots\ be shown that the quantum mechanical correlations cannot be reproduced 
by a hidden variable theory even if one allows a ``local" sort of indeterminism. For example, one could imagine that the indeterminism might be introduced at every space-time point and allowing the result of each throw to influence physical events in the future light cone of the point in question. This would not work: the quantum mechanical correlations are too perfect to permit any such statistical slop. Popper remarks \{in a volume published in 1971\} that he does not find this point manifest in my own paper on the subject, but it is there --- very briefly.
\end{quote}
Here Bell's ``own paper" is Ref.~\cite{Bel64}, and the implication is that Bell believed he covered 
the \guillemotleft``local" sort of indeterminism\guillemotright\ in that paper. Presumably the ``very brief" 
manifestation Bell has in mind 
is his  argument  from locality as per Definition~\ref{def:local} to determinism 
via predictability --- the incorrect argument I discussed in Sec.~\ref{sec:Bell64}. 

In the concluding paragraph of his 1976 paper, Bell writes:
\begin{quote}
This paper has been an attempt to be rather explicit and general about the notion of locality, along lines hinted at in previous publications \{\cite{Bel64,Bel71,Bel72,Bel73}\}.
\end{quote} 
 With regard to the later references, this quote is more than fair, as 
the ``certain local character''  of Ref.~\cite{Bel71} did capture the implications of 
the notion of LC, as discussed in Sec.~\ref{sec:prec}.  
With regard to Bell's 1964 paper \cite{Bel64}, it is also true that the explicit notion of locality (L) 
used there was not as {\em general} as LC. But the latter was {\em hinted at} only in so far as: 
(i) it is related to (though much more general than; see \ref{App:EPR}) 
the notions used by EPR, to whom Bell attributes his pseudoderivation of D from predictability; 
(ii) {\em unlike} Bell's 1964 explicit notion (L),  it {\em does} allow one to derive D from predictability.   
Point (ii) is easy to show: If $P(A|a,B,b,c) = 1$ or $0$, then the same must be true for all 
$P_\theta(A|a,B,b,c,\lambda)$.\footnote{Strictly, this need not hold 
for those $\lambda$ such that $d\mu_\theta(\lambda|c) = 0$, that is, for those $\lambda$ which 
never occur in the experiment. We are not concerned with such $\lambda$ in definitions of D, L, or LC.}  
But if we assume LC (Criterion~\ref{def:LC}) it follows that 
$P_\theta(A|a,c,\lambda) = 1$ or $0$ also, which is L\&D. 

In any case, it seems that once Bell had explicitly defined LC, he wished all previous localistic 
notions he had used,  
in particular the notion of locality as per Definition~\ref{def:local},  to be forgotten. 
Moreover,  after a few years he became convinced   that it was the notion of LC that he had in mind all along. 
In 1981 Bell says (Ref.~\cite{Bel81}, text and footnote 10 stitched together) 
\begin{quote}
It is important to note that to the limited degree to which {\em determinism} plays a role in the EPR argument, 
it is not assumed but {\em inferred}. What is held sacred is the principle of ``local causality" --- or ``no action at a distance". \ldots\ There is a widespread and erroneous conviction that for Einstein (and his followers) determinism was always the sacred principle. (My own first paper on this subject~\{\cite{Bel64}\} starts with a summary of the EPR argument from locality to deterministic hidden variables. But the commentators have almost universally reported that it begins with deterministic hidden variables.) 
\end{quote}
Thus, Bell implies in 1981 that both he and Einstein were always using the notion of LC, 
which Bell characterises later in this 1981 paper in the same way (Criterion~\ref{def:LC}) as in 1976.  
As argued in Sec.~\ref{sec:Bell64},  there is only one plausible reading of ``locality'' 
in Bell's 1964 paper, and it is not LC. Nor  did EPR formulate such a general notion, 
 Refs.~\cite{DGZ92,Nor06} notwithstanding. 

Let this be absolutely clear: I am not accusing Bell or his followers of intellectual dishonesty. Local 
causality is such an elegant notion that in hindsight it is astonishing that it had to wait until 1976 
(or at least 1974~\cite{CH74}) 
to be  applied to quantum phenomena.  If EPR had stated this assumption in 1935, they could have used their 
example to formulate a far simpler argument showing OQM  to be an unsatisfactory theory  (by their lights).  
Indeed, Bell credits them with exactly this argument, in Sec.~6.8 of 
Ref.~\cite{Bel90b}\footnote{Of course EPR would not necessarily have endorsed the name \red local causality \blk  
in this context 
because their aim was to prove the incompleteness of OQM. That is why they built localistic 
notions into their definition of completeness; see \ref{App:EPR:Com}.}. 
Furthermore LC seems to me a more natural localistic notion than L, 
as will be discussed in Sec.~\ref{sec:Disc}. The fact that Bell thought, mistakenly, that  one \red could \blk obtain   
determinism from locality (as defined by him in 1964) suggests strongly that he already had then an intuitive 
localistic notion (\eg~``no action at a distance" \cite{Bel81}) that was different from, and stronger than, 
L. From personal experience I know that it is easy to forget, after a decade or two, the details of a 
published argument, 
and to remember the intuitive idea which in fact only crystalised in one's mind later. 
I have no doubt that anyone familiar with Bell's later work could have educed from Bell in 1964 
the precise notion of LC in Criterion~\ref{def:LC}, with little effort on either's part.  But 
hypotheticals do not alter history, and  the theorem  Bell {\em proved} in his 1964 paper is not the same as 
Theorem~\ref{thm:Bell76} from 1976.

\section{Related Theorems} \label{sec:related}
 
The interpretation of Bell's theorem has generated a large literature. In this section I briefly review 
some of the results that will be  relevant  to my Discussion section. 

\subsection{Fine}

As discussed above, Clauser and Horne in 1974 \cite{CH74} recognised that factorisability (\ref{factor}) 
--- which is  a consequence of  LC --- is no more general than Bell's 1971 ``certain local character" \cite{Bel71}. 
And in that paper Bell noted that the same formula (\ref{certainlocal}) applied to correlations under 
this constraint as under the constraint of L\&D. That is, 
\begin{theorem}[Fine 1982] A phenomenon respects LC iff it respects L\&D. 
\label{thm:Fine}
\end{theorem} \gold
Note that this theorem is not saying that LC and L\&D are logically equivalent concepts. Recall that 
there are theories that satisfy LC but that do not satisfy L\&D, such as OQM for product states.

I call the above  Fine's theorem, even though his paper~\cite{Fin82} came much later than the above, because 
it was he who focussed attention on it, rather than burying it in footnotes, as in Refs.~\cite{CH74,Bel71}. 
 Fine refers to  ``factorizable stochastic models'' (\ie\ F of Definition~\ref{def:factor}) 
rather than to LC, but in terms of  the observable events they are equivalent.  
Fine duly credits Clauser and Horne \cite{CH74} with Theorem~\ref{thm:Fine}, but also establishes it directly, 
for the case of two settings per party, each with two outcomes \cite{Fin82}. 
The `if' part of Theorem~\ref{thm:Fine} is trivial, since L\&D implies F. But F does not imply L\&D, so the non-triviality is in the `only if' part. For a general proof, see Appendix A of Ref.~\cite{Hal11}. The basic idea, as in Ref.~\cite{Bel71}, is that in a  factorizable  theory, any stochasticity \red in an outcome \blk can be {\em reproduced} by \red having it determined by 
additional \blk hidden variables that may be assumed to reside locally.  
\subsection{Jarrett / Shimony} \label{sec:JarShi}

Fine's theorem showed that, in terms of its experimental consequences, LC might as well be thought of 
as L\&D. Not long after, Jarrett \cite{Jar84}  showed 
that LC is in fact logically equivalent to a subtly different conjunction. He (and, independently, 
Shimony~\cite{Shi84}) introduced a new concept,
\begin{definition}[Jarrett-completeness]
A theory satisfies Jarrett-completeness ({\rm JC}) iff	
	\beq
	P_\theta(B|A, a, b, c, \lambda) = P_\theta(B|a, b, c, \lambda).
	\eeq
\end{definition}
Using this he proved 
\begin{theorem}[Jarrett 1984] \label{thm:Jarrett}
 Factorizability  (or ``strong locality" \cite{Jar84}) is equivalent to the conjunction of: 
\begin{enumerate}
	\item Locality\footnote{Jarrett \cite{Jar84} agrees with Definition~\ref{def:local}. So does Howard~\cite{How85}. But the latter uses the term ``separability" for Jarrett-completeness, a usage which for me is not intuitive, and which has not caught on. Recall that Bell used the terms ``separable" and ``local" interchangeably in his earliest papers. One might be tempted to call Jarrett-completeness ``causality" because of Theorem~\ref{thm:Jarrett} (since factorisability amounts to local causality here) but it is clear that these concepts are already overloaded with proposed names.}, also known as ``parameter independence" \cite{Shi84}.
	\item Jarrett-completeness, also known as  ``outcome independence" \cite{Shi84}.	\end{enumerate}
\label{JarShi}
\end{theorem}
The corollary, via Theorem~\ref{thm:Bell76}, is
\begin{corollary}[Bell 1976 as L\&JC] \label{thm:BellLJC}
Quantum phenomena violate either locality, or Jarrett-completeness. 
\end{corollary}  
Here my theorem-naming convention indicates that Bell's 1976 argument would have gone through with the 
assumptions of L\&JC rather than the assumption he did make there (LC); the same 
convention applies to Theorems~\ref{thm:BellFLJC}, \ref{thm:BellFLD}, and \ref{thm:Bell1} below.  

Orthodox quantum mechanics of course satisfies L but violates JC,  while 
deterministic hidden-variable theories (such as Bohm's theory \cite{Boh52}) satisfy JC but violate L. 
Thus the options presented by Theorem~\ref{thm:BellLJC} (Jarrett's version of Bell 1976) are resolved 
in the same way as in Theorem~\ref{thm:Bell64} (Bell 1964) for these two leading approaches to 
QM (of those that satisfy the axioms of Sec.~\ref{sec:Assumptions}). 
This is not surprising because JC can be thought of as a weak version of determinism: 
\begin{lemma}
Determinism implies Jarrett-completeness.
\end{lemma} 
Shimony \cite{Shi84} called the violation of L ``controllable nonlocality", which is perhaps fair enough. 
However it is another matter to state that such violations ``provide (at least in principle) the means for superluminal signal transmission," as does Jarrett~\cite{Jar84} (and, \red in much the same words, \blk also Shimony~\cite{Shi00}). 
For example, although Bohm's theory violates locality, its statistical predictions are identical to OQM, and so it does not allow superluminal signalling.  This is because the value of a particle's position (a hidden variable in Bohm's theory) 
cannot be known to a macroscopic agent any  better than in OQM. In some approaches to Bohm's theory~\cite{Wis07} --- 
although not others~\cite{Val02} --- this lack of knowledge is inevitable. 
It is only for {\em operational} theories that 
locality is the same as signal-locality \cite{Sky82}, 
\begin{definition}[signal-locality] \label{def:sl}
A phenomenon respects signal-locality  iff $f(B|a,b,c) = f(B|b,c)$.
\end{definition}
See Ref.~\cite{Dic98} for further discussion of the distinction between L and signal-locality.

It is important to note, again {\em contra} Jarrett \cite{Jar84}, that only signal-locality is an absolute 
``requirement of relativity theory",  which can be seen as follows. Let us assume, in addition to 
Axioms~\ref{axiom:macro}--\ref{axiom:willy}, Principle~\ref{pr:efficacy}, for  manifest  \red agent-causation\blk. 
Now since Definition~\ref{def:sl} involves relative frequencies, any violation of 
signal-locality would mean, by Principle~\ref{pr:efficacy}, that $a$ is a cause for $B$. But because 
$a$ is not in the past light-cone for $B$, Principle~\ref{pr:relativity} (\red relativistic causality\blk),  would be 
violated. By contrast, if a {\em theory} violates L, but not signal-locality, these assumptions do not 
entail that $a$ causes $B$, because the {\em phenomenon} exhibits no dependence of $B$ on $a$. 
Hence there is no conflict with \red relativistic causality\blk. I will revisit this point in Sec.~\ref{sec:irrec}. 
 

Jarrett naturally calls JC merely ``completeness"; I have \red antepended \blk his name 
because this concept is neither implied by, nor implies, EPR-completeness,  
$\Com(\theta)$, as I have analysed it (see \ref{App:EPR:Com}). 
In Jarrett's defence, however, one can relate them with only a weak additional assumption 
as follows. First I will define 
\begin{definition}[\red fragile locality] \label{def:fragile-locality}
A theory satisfies {\red fragile \blk locality} ({\rm {\red FL}}) iff 
\beq 
P_\theta(B|a, b, c, \lambda) \in \{0,1\} \implies P_\theta(B|a, b, c, \lambda) = P_\theta(B|b, c, \lambda) .
\eeq
\end{definition}
That is,  if, for some values of its arguments, there exists a function 
$B_\theta(a,b,c,\lambda)$, then that  function does not depend upon $a$. 
\red The terminology ``fragile" is appropriate because it depends upon exact  
determination, which is a set of measure zero in correlation-space. 
Note that \blk Jarrett's ``strong locality" (\ie\ \red factorizabiltiy, or, more or less, \blk LC) 
implies locality, which in turn implies \red fragile \blk locality.  
 
Given Definition~\ref{def:fragile-locality},  it is not difficult to show (see \ref{App:EPR:Jar}) that 
\begin{theorem} \label{thm:JCFLEC}
If EPR-completeness implies X, then Jarrett-completeness plus \red fragile \blk locality implies X.
\end{theorem}
(Indeed, it could be argued that \red fragile \blk locality is one of the localistic notions that EPR implicitly assume; 
see \ref{App:EPR:Com}.) A corollary of this, via Theorem~\ref{thm:Bell1}, is 
\begin{corollary}[Bell 1964 as {\red FL}\&JC]\label{thm:BellFLJC}
Quantum phenomena violate  either \red fragile \blk locality, or Jarrett-completeness. 
\end{corollary} 
 Note that the joint assumption of \red fragile \blk locality and Jarret-completeness 
does {\em not} imply LC. Thus, this version (Corollary~\ref{thm:BellFLJC}) 
of Bell's 1964 theorem, unlike Bell's original version,  
does not have stronger premises than Bell's 1976 theorem. That is, it is not 
subsumed by Theorem~\ref{thm:Bell76} (using LC) in the way that Theorem~\ref{thm:Bell64} 
(using L\&D) is. The same remarks also apply to Theorem~\ref{thm:Bell1}. 

Like Theorem~\ref{thm:Bell1}, the statement of Bell's 1964 theorem 
as Corollary~\ref{thm:BellFLJC}  
can only be proven for quantum phenomena with perfect correlations (as Bell considered in 1964). 
Thus it does not say anything about phenomena that are experimentally accessible; 
see \ref{App:EPR:Bel} for details. 
This  shortcoming   in the theorem can be removed by strengthening  {\red FL} to L 
(giving Theorem~\ref{thm:BellLJC}), or JC to D, giving 
\begin{corollary}[Bell 1964 as {\red FL}\&D] \label{thm:BellFLD}
Quantum phenomena violate either \red fragile \blk locality, or determinism. 
\end{corollary} 
I call this a version of Bell's 1964 theorem because D\&{\red FL} $\iff$ D\&L 
(even though only L $\implies$ {\red FL}). 
In fact one might deem Bell's characterisation of locality in Ref.~\cite{Bel64} 
to be too imprecise to distinguish between L of Definition~\ref{def:local} and 
{\red FL} of Definition~\ref{def:fragile-locality}, but to avoid further terminological 
complication I will not pursue that line of argument.

\section{The Two Camps} \label{sec:2camps}

As stated in the Introduction, those who ponder the implications of Bell's theorem, 
(physicists, philosophers, and, increasingly, information scientists), 
and who accept Axioms~\ref{axiom:macro}--\ref{axiom:willy}, can broadly be grouped into two 
camps: operationalists and realists. Note that these \red four Axioms respectively \blk rule out: 
relativist approaches such as Everett's relative state interpretation \cite{Eve57} (popularly 
known as the many worlds interpretation \cite{DeWGra73}); 
giving up free choice (albeit maybe to only a modest degree~\cite{Hal10}), 
which Bell called ``superdeterminism''~\cite{Bel90b}; ubiquitous 
wormholes, \red as discussed in Ref.~\cite{Holland95} (pp.~481--3); \blk and retrocausal ideas \cite{Peg80,Price08}.

\subsection{Operationalists}

The operationalist attitude is particularly common among quantum information theorists, and 
is well represented by this quote from a recent text-book \cite{SchWes10}:
\begin{quotation} 
\{To derive Bell's theorem, we\} now make two hypotheses about the behaviour of the composite system: 
\begin{description}
\item[Hidden variables.] We assume that the results of any measurement on an individual system are predetermined. Any probabilities we may use to describe the system merely reflect our ignorance of these hidden definite values, which may vary from one experimental run to another.
\item[Locality.] We assume that Alice's choice of measurement does not affect the outcomes of Bob's measurements, and {\em vice versa}.
\end{description}
As we have seen, these two hypotheses more or less capture the point of view advocated by EPR.
\end{quotation}
Note that Schumacher and Westmoreland use exactly 
the same terminology ({\em hidden variables}, {\em predetermined}, and 
{\em locality})  as Bell's 1964 paper, and I think their presentation very fairly captures the definitions in that paper. 
As for its implication, they state
\begin{quote}
From a philosophical point of view, Bell's theorem and the experimental confirmation of quantum theory constitute one of the most remarkable results in all of physics. The dual hypotheses of hidden variables and locality have a pretty good claim to the ``common sense" view of the world. Nevertheless, {\em at least one of them must be wrong.} 
\end{quote}

Another famous quantum information text-book with a similar presentation is Ref.~\cite{Nielsen00a}. 
Showing some appreciation of the literature, they say 
\begin{quote}
Vast tomes have been written analyzing 
\ldots\ the subtly different assumptions which must be made to reach Bell-like inequalities. Here we merely summarise the main points.
\end{quote}
But their summary is very similar to that of Ref.~\cite{SchWes10}. 
In place of ``hidden variables" or ``predeterminism" they state 
\begin{quote}
The assumption that \ldots\ physical properties \ldots\ have definite values \ldots\ which exist independently of observation. This is sometimes known as the assumption of {\em realism}. 
\end{quote}
Having then stated the locality assumption (practically identically to Ref.~\cite{SchWes10}), they go on:
\begin{quote}
These two assumptions together are known as the assumption of {\em local realism}. They are certainly intuitively plausible assumptions about how the world works, and they fit our everyday experience. Yet the Bell inequalities show that at least one of these assumptions must be incorrect. 
\end{quote}
 
 It is worth making a few remarks about this terminology. ``Local realism" is certainly a common term these days 
 for what Bell experiments disprove. As far as I can tell, it was introduced by d'Espagnat in 1979 \cite{dEs79}, 
 although it is reminiscent of the term ``objective local" used by Clauser and Horne \cite{CH74}. 
 However, the term ``realism" was never used by Bell, and is generally used synonymously with 
  (pre)determinism, as Mermin (another early proponent) admits~\cite{Mer80}. It is because of this 
  that I have 
  called Axiom~\ref{axiom:macro} --- the independent existence of detectors and results ---  ``macroreality". (For further discussion of the infelicity of ``local realism", see Ref.~\cite{Nor07}.) By my casual observation, ``realism" is used 
  more often by those who see it as the obvious hypothesis to reject, such as Nielsen and Chuang \cite{Nielsen00a}:  
  \begin{quote}
Most physicists take the point of view that it is realism that must be dropped from our worldview in quantum mechanics.
\end{quote}

Because  operational QM xsatisfies locality, it is natural for operationalists to advocate espousing locality, and rejecting the second assumption required for Bell's 1964 theorem. Nevertheless I find it a surprising development in the language surrounding Bell's theorem, that this second assumption should often be called ``realism". 
The usual philosophical meaning of ``realism" is the belief that entities exist independent of the mind, a worldview one might expect to be foundational for scientists. However, there is an influential  sub-community of operationalists,  
the most vocal  among  which go by the name of QBists (Quantum Bayesianists), 
who reject precisely that, saying \cite{FucMerSch14} ``reality differs from one agent to another."  
For QBists, reality \red is identified with \blk the experience of a single agent, and they dismiss Bell's theorem by 
emphatically rejecting Axiom~\ref{axiom:macro} (macroreality).  Each QBist believes in his own reality, but refuses to 
\begin{quote}
 assign correlations, spooky or otherwise, to space-like separated events, since they cannot be experienced by any single agent. Quantum mechanics is thus explicitly local in the QBist interpretation.
\end{quote}
As Louis XIV might have said, \guillemotleft L'\'etat quantique, c'est moi.\guillemotright

\subsection{Realists} \label{sec:realists}

Since operationalism, taken to its logical extreme in QBism, is anti-realist in all senses of the word, 
it seems appropriate to use 
the term ``realists" for those at the other end of the spectrum in the debate on the meaning of 
Bell's theorems. This camp, in which we can locate Bell himself,  even if he was never so dogmatic 
as some~\cite{DGZ92,Mau94,Nor06}, is characterised by two convictions: 
(1) that 
correlations need to be explained (Bell~\cite{Bel81} calls this nothing less than ``the scientific attitude''); 
(2) that nature should have a unified description, in which anthropocentric notions such as 
``detector settings" should play no fundamental role. 

Despite the name I have given it, the realist camp does not 
advocate that we should cleave to ``reality''~\cite{Nielsen00a,Mer80} (that is, determinism, also expressed as 
``hidden variables''~\cite{SchWes10,Mermin93}) and 
spurn the locality of Definition~\ref{def:local}. Rather, its members generally read Bell's theorem as requiring 
(in addition to the axioms of Sec.~\ref{sec:Assumptions}) only a single assumption \cite{DGZ92,Mau94,Nor06}, and 
 often explicitly reject the two-assumption formulation of the operationalist camp. 
A quote from the textbook of Maudlin \cite{Mau94} (p.~19) will suffice:
\begin{quotation}
Bell himself derived the result as part of an examination of so-called local hidden-variable's theories \{which add\} 
parameters whose values determine the results of experiments. Bell's results are therefore sometimes portrayed as a proof that local deterministic hidden variables theories are not possible. 

This is a misleading claim. It suggests that the violation of the inequality may be recovered if one just gives up determinism or hidden variables. But as we have seen, the only assumption needed to derive the inequality is that the result of observing one particle is unaffected by the experiment carried out on the other \ldots\ a condition \{which\} is generally called ``locality". 
\end{quotation}

At first sight this seems in blatant conflict with the operationalist camp who claim that an additional assumption to locality, which I would call determinism, {\em is} necessary to derive Bell inequalities. But on closer inspection, Maudlin's definition of locality is subtly different. It refers not to the effect of ``Alice's choice of measurement" \cite{SchWes10}, 
but to Alice's ``experiment" (or, a few pages earlier, Alice's ``observation") 
which presumably includes the outcome she gets as well as her choice of measurement.  
That is, by ``locality", Maudlin means ``local causality" 
as introduced by Bell in 1976.

The preference for the 1976 theorem was Bell's own, as already noted, and 
the Conclusion to his final paper on the matter expresses the realist view well~\cite{Bel90b}:
\begin{quotation}
The obvious definition of ``local causality" does not work in quantum mechanics, and this cannot be attributed to the ``incompleteness" of that theory. \ldots\

Do we then have to fall back on ``no signalling faster than light" as the expression of the fundamental causal structure of contemporary theoretical physics? That is hard for me to accept. For one thing we have lost the idea that {correlations can be explained} \ldots. More importantly, the ``no signalling \{faster than light\}" notion rests on concepts which are desperately vague, or vaguely applicable. The assertion that ``we cannot signal faster than light" immediately provokes the question 
	\begin{quote}
	Who do we think {\em we} are?
	\end{quote}
\end{quotation}  
Bell clearly expresses the two tenets of the realist camp, as listed in the first paragraph of this section, 
but here gives greater weight to the second. 
While ``no signalling faster than light" is not the same as locality (see Sec.~\ref{sec:JarShi}), the objection to using anthropocentric concepts (\ie~actions by agents) as the basis for physical theory applies equally to both. 

 The important point is that the notion of LC does not rely on action by agents: a theory could violate 
LC with no agent choice at all. For example, consider a single quantum particle in a pure state split between two 
boxes, one sent to Alice and the other to Bob. Then simultaneously (by some foliation) they open the boxes,  
and observe whether the particle is there or not. Clearly their observations will be perfectly correlated 
(the particle is seen by one and only one of them) but in OQM there is no variable $\lambda$ that 
factorizes the joint probability distribution as per \erf{factor}; there is no variable that determines 
which box the particle will be found in. Thus OQM can be shown to violate LC 
without measurement choice. (This is sometimes known as the ``Einstein Boxes'' argument \cite{Nor05}.)
The subtlety of Bell's insight is that the only {\em phenomena} that violate LC 
 {\em do} require agency (free choice of settings).  


\subsection{Irreconcilable differences?} \label{sec:irrec}

If one has ambitions to reconcile these two camps, then first one must understand the reasons for 
the differences between their presentations.

 The first, and perhaps most  important, difference is in terminology, as I have already discussed. 
As I have shown, exemplars in Bell's writings exist for both, competing, definitions of ``locality". 
The operationalist camp use it as Bell used it 1964 and a few subsequent papers, and the realist 
camp as he used it, occasionally, from 1976 onwards, as a synonym for LC. 
The second difference is that the two camps then, as expected, state Bell's theorem as 
Theorems~\ref{thm:Bell64} and \ref{thm:Bell76} respectively. 

Being followers of Bell, the realist camp must be aware that, from 1976 onwards, 
Bell used the term ``local causality" in preference to ``locality", especially in formal 
expositions. While some prominent realists sometimes promote Bell's preferred term \cite{Nor11}, 
it is more common to see ``locality" used for LC. For example, the above quote from Maudlin's 1994 book \cite{Mau94} is unchanged in the 2011 edition, while I must admit that in an earlier work \cite{Wis06a} I mentioned ``local causality" only to say that it was ``often abbreviated to locality," which I thereafter used. 
Presumably this  substitution is made  because ``locality" is (a) shorter 
and thus more memorable, (b) a single word and thus not able to be confused with a dual concept; and (c) 
simpler and thus suggestive of a natural concept. My present stance is that these advantages do not outweigh  
the disadvantages: (1) it disregards Bell's final and most thorough communication on 
the subject~\cite{Bel90b}; and (2) it makes it appear as if the two camps are in disagreement over
basic logic, when in fact that is not the case. 

The realist camp might object along these lines: \guillemotleft Why should we 
change our terminology? Let the operationalist camp use the phrases ``parameter independence" \cite{Shi84} 
or ``setting independence" if they need a term for such an unnatural notion.\guillemotright\ 
There are two answers to that. First, the operationalists (being, by nature, primarily interested in 
the applications of quantum mechanics) are generally less well read in this area of metaphysics, 
and are less likely to be familiar with such alternate phrases. Second, the one piece 
of literature everybody reads (or at least references) with regards Bell's theorem is his 1964 paper, 
and there the operationalist (or indeed anyone without the ``benefit" of hindsight) will read  
Definition~\ref{def:local} for ``locality" in Bell's words no fewer than four times, as well 
as a similar concept in Einstein's words.
 
For the operationalist, locality is a natural assumption because OQM respects it,   
and because it follows from the assumptions of Sec.~\ref{sec:Assumptions}, {\em if} 
one strengthens Principle~\ref{pr:efficacy} to the (still reasonable) 
\begin{principle}[\red agent-causation\blk] \label{pr:ce}
If, in a theory, an event is statistically dependent on a freely chosen action, 
then, in that theory, that action is a cause of that event.
\end{principle}
Holding to this principle  would be,  as far as I can tell, the  simplest  
way for the operationalist camp to justify L from relativity (Principle~\ref{pr:relativity}), 
and thereby conclude, from Theorem~\ref{thm:Bell64} (Bell 1964), that D must be 
violated\footnote{Alternatively, since signal-locality is a matter of fact about quantum phenomena 
(and the world, as far as we know), it would presumably  be possible to derive L using 
some sort of ``no fine-tuning'' Principle (\ie~that the causal structure of the theory should 
not change under small changes in the probability distribution for the hidden variables). 
There is a similar argument deriving LC from signal-locality, 
Reichenbach's Principle~\ref{pr:cc}, and ``no fine-tuning''~\cite{WooSpe12}.}.
 While realists would not necessarily reject Principle~\ref{pr:ce}, they would not see it as 
an appropriate one for fundamental analysis of the physical world, because 
of the central role it gives to ``freely chosen action'' (see the extended quote from 
Bell in the preceding section).   

The challenge for the the operationalist camp is 
to recognise that there is another localist notion (I speak of LC of course)  
which OQM does {\em not} respect, and which is not only meaningful,  but which 
also follows from the assumptions of Sec.~\ref{sec:Assumptions} (with or without 
Principle~\ref{pr:efficacy}), {\em if} one holds to another reasonable principle: 
\begin{principle}[common cause \cite{Rei56}] \label{pr:cc}
If two events with no direct causal relation are correlated,  then they have a common cause $\kappa$, such that conditioning on $\kappa$ will eliminate the correlation. This is so even if the correlation is already conditional, 
as long as the conditioning \red event is not an effect of both events.\footnote{\red With regard to the last clause, Axiom~\ref{axiom:arrow} implies that 
as long as an event (say $f$) is in the past light cone of event $e$, it cannot be an effect of $e$.
Moreover, if Principle~\ref{pr:relativity} is also 
assumed, then unless $f$ is within the future light cone of $e$, it cannot be an effect of $e$. \blk } \blk
\end{principle}
Just as local causality $\implies$ locality $\implies$ signal locality, 
we have for these principles 
\begin{theorem}
common cause $\implies$ \red agent-causation \blk $\implies$ 
manifest \red agent-causation\blk. 
\end{theorem}
The less obvious (the first) of these implications can be seen as follows. If an event $A$ is 
statistically dependent upon a free choice $a$, that means that they are correlated. Now $a$ is uncaused by Axiom~\ref{axiom:willy}, 
so by Principle~\ref{pr:cc} the only option is that $a$ causes $A$, as in Principle~\ref{pr:ce}.

For the realist camp, Principle~\ref{pr:cc} is  fundamental (correlations need to be explained), 
and is used implicitly  to motivate LC (rather than L) from 
relativity.  Thus,   since Theorem~\ref{thm:Bell76} (Bell 1976) rules out LC,
 members of the realist camp would 
typically reject Principle~\ref{pr:relativity}. That is, they would allow space-like causal influences, 
as long as they respect Axiom~\ref{axiom:arrow} according to some foliation of 
space-time. 
 As an aside,  it seems to me that, for this purpose,  
one could use, in place of Principle~\ref{pr:cc}, the rather simpler
\begin{principle}[\red causal completeness\blk] \label{pr:gc}
 The probability of an event, conditioned on its causes, is unchanged if further conditioned 
on other events \red that are not its effects. \blk
\end{principle}

 To close, the above   challenge to operationalists evokes  a third reason the realist camp should 
 not try to own ``locality": (3) by using ``local causality" and referencing Bell's 1976 paper, realists might 
induce operationalists to read something else by Bell, and to appreciate the power
of Criterion~\ref{def:LC} and Theorem~\ref{thm:Bell76}.  That said, operationalists  
would still not be compelled to accept space-like causal influences, as they could follow Ref.~\cite{vFr82} 
and reject Principle~\ref{pr:cc} (common cause)  rather than Principle~\ref{pr:relativity} (\red relativistic causality\blk). 
This is simply a sophisticated way of saying that operationalists could still, in Bell's words \cite{Bel80},   
``shrug off a correlation.'' 
 
\section{Discussion} \label{sec:Disc}

Bell's theorem  is the  most profound ramification of quantum theory  that has been experimentally confirmed.    
In this paper I have presented several different formulations of it: 
Theorems~\ref{thm:Bell64}, \ref{thm:Bell76}, \ref{thm:BellLJC}, \ref{thm:BellFLJC}, 
\ref{thm:BellFLD}, and \ref{thm:Bell1}. However, I would argue that there are only two essential forms with which 
quantum physicists  should  be familiar, corresponding to the two Bell's theorems 
of my title, Theorem~\ref{thm:Bell64} (Bell 1964), and Theorem~\ref{thm:Bell76} (Bell 1976).
Even though either theorem implies the other, the two have independent roles in quantum foundations, 
as the expressions of Bell's insight preferred by operationalists 
and realists respectively.

Of these I side with the realists in thinking that Bell's 1976 theorem is the  more  important. 
The  assumption of local causality  is strictly weaker than the assumption of locality 
and determinism, so Bell's 1976 theorem subsumes Bell's 1964 theorem. Also, the notion 
of local causality  seems to me 
the most natural expression of the spirit of special relativity for statistical theories: 
that, given a complete (within the theory) specification of the  physical situation  
 prior to Alice's and Bob's choices of measurement settings,   
Alice's outcomes are statistically independent of space-like separated events (such as Bob's 
choice and his outcomes). Then, as Bell proved,  
``quantum mechanics is not locally causal and cannot be embedded in a locally causal theory''~\cite{Bel76}.
That is, orthodox quantum mechanics violates local causality, and 
adding hidden variables cannot alleviate the problem.

I have argued strongly (in Secs.~\ref{sec:Bel76} and \ref{sec:irrec}) 
for using the term ``local causality", not ``locality", in this statement of 
Bell's theorem for two principal reasons. First, it follows the formal definition Bell proposed in 1976~\cite{Bel76}, 
and the usage he \red resolutely \blk promoted in his ultimate presentation~\cite{Bel90b}. Second, and more importantly, 
it is necessary to avoid the confusion (and even hostility) which can arise when people read Bell's 1964 
paper --- or other seminal papers such as Refs.~\cite{CHSH69,Aspect81,Aspect82,Jar84} --- and see there 
``locality" defined in a weaker (\ie~less general) sense, in which 
Alice's outcomes are required only to be statistically independent of Bob's 
choice of measurement settings.  Another reason to endorse the terminology 
``local {\em causality}" is that, under Axioms~\ref{axiom:macro}--\ref{axiom:willy}  and 
\red relativistic causality \blk (Principle~\ref{pr:relativity}),  
local causality is simply the expression of Reichenbach's ``Principle of common cause''~\cite{Rei56}, 
Principle~\ref{pr:cc}.

If (as I am advocating) Bell's promulgation of the term ``local causality" should be respected, 
then why not also his abandonment of his earlier notion of locality?   Unfortunately for Bell, 
one cannot choose which contribution one will be remembered for. 
Bell's 1964 paper is, and probably always will be, the most famous of his papers, the one 
to which new researchers in the field will turn as a primary source. It is thus hard to argue that 
 ``locality" as per Definition~\ref{def:local} should not be taken seriously as a concept when 
 it played such a foundational role.  
 Also, locality  can be derived from  assumptions~\ref{axiom:macro}--\ref{pr:relativity},  together with the 
reasonable  Principle~\ref{pr:ce} which I have tentatively called  ``\red agent-causation\blk''. 

Bell's 1964 theorem --- that quantum phenomena are incompatible with either 
locality or determinism --- is naturally favoured by operationalists because operational 
quantum mechanics {\em does} respect locality, leaving determinism as the obvious 
assumption to forgo. Rather than determinism (or ``hidden variables" or ``realism"), 
some have suggested a weaker assumption (see Sec.~\ref{sec:JarShi}) 
which can be combined with locality  in order to derive a contradiction with 
the predictions of quantum mechanics (Theorem~\ref{thm:BellLJC}). I do not 
believe such formulations shed much light beyond Bell's original formulation, any 
more than I believe Theorems~\ref{thm:BellFLD} (replacing locality with a weaker assumption) 
or~\ref{thm:BellFLJC} (replacing both locality and determinism with weaker assumptions) do. 

 For operationalists, Bell's 1964 theorem is not only more natural, it is also 
more useful, in the following sense. As an operationalist, one is interested only in theories in which 
the hidden variables ($\lambda$) are, in principle, knowable to some agent. Such theories must be local 
if signal locality is to hold even when the knowledge of all agents
is taken into account. Similarly, such theories must be deterministic if the measurement 
outcomes are to be predictable when taking into account the knowledge of all agents. 
Thus, if one believes in signal locality as a matter of principle, then Bell's 1964 theorem implies that 
quantum phenomena must be unpredictable in principle~\cite{Mas06}. (See Ref.~\cite{CavWis12} for a 
more formal presentation.) 
In information science, guaranteed unpredictability 
or randomness is a valued resource for numerical algorithms and security protocols~\cite{Pir10}. 
 It is thus of interest that 
a Bell experiment can be turned into a protocol that generates (or, rather, expands) 
randomness, certified by the principle of signal-locality~\cite{Pir10}. 
 Moreover, because Alice's and Bob's results 
are {\em correlated} in Bell-type experiments, it is possible, in principle, to 
use them to obtain {\em secret} {\em  shared randomness} between distant parties, 
which is even more valuable, enabling communication with security guaranteed by 
signal locality~\cite{Bar05}\footnote{In fact, other assumptions are also necessary to guarantee security, 
in particular that neither Alice's nor Bob's laboratory inadvertently leaks any kind of information. The proof in 
Ref.~\cite{Bar05} requires perfect correlations (as do the protocols in the subsequent paragraph), but the idea can be applied in quantum key distribution 
even with imperfect correlations, as follows. If one assumes the correctness of quantum mechanics (a strictly stronger assumption than signal locality), then violating a Bell inequality may allow the distribution of a secret key without Alice or Bob having to trust the entanglement source or their detectors~\cite{Acin07}. Similar results apply for randomness expansion~\cite{Pir10}.}.

While distributing a secret key is the most important proposed application of Bell's inequalities, 
 there are more specialised applications of quantum correlations that are, in themselves, a proof of  
 violation of local causality, without statements about secrecy being required. In particular, there are  \azur
 protocols  --- such as those enabling teams of accused robbers~\cite{JacWis05} or 
 crooked accountants~\cite{Sud07} to lie to their interrogators without fear of discovery --- 
 in which {\em all} correlations are perfect\footnote{The class of quantum phenomena which violate local causality in this way has been called ``quantum pseudo-telepathy''~\cite{Bra04}, an echo of Einstein's term (see~Supposition~\ref{sup:NT} and the  subsequent discussion).}, and the appearance of predetermination  
 is so strong that it is hard not to feel that  they must be nonlocal.  Nevertheless, 
 according  to the terminology  I recommend,  neither of Bell's 
theorems allow us to conclude that quantum phenomena are nonlocal,  taking that word to mean  
the negation of ``local".

 In conclusion, for a proper appreciation of the foundational importance 
of Bell's theorem to physics, information science, and the philosophy of causation, one should 
be familiar with both the 1964 Bell's theorem and the 1976 Bell's theorem, even though they 
are logically equivalent. The former proves that quantum phenomena are either nonlocal  
(in a ``\red causation by agents\blk'' sense) or undetermined, while the latter proves that quantum phenomena 
violate local causality (in a ``common cause for correlations'' sense). For those who 
prefer the latter theorem, as Bell ultimately did, and who find the verb phrase ``violate local 
causality'' lacking in pith,  I make one final terminological 
suggestion,  regarding a pre-existing term: 
\begin{definition}\label{def:BN} 
`Bell-local' is an acceptable synonym for `locally causal', and `Bell-nonlocal' for its negation.
\end{definition} 
This usage has a long history  (going back at least to 1981~\cite{dEs81},  in the negative) and is still 
very much current (see \eg~the review~\cite{Bru14}).
 It is not a term which Bell would ever have used, 
 and, because of Bell's inconsistent use of ``locality",  it risks confusing unwary 
readers of the `historical' literature on Bell-nonlocality\footnote{For 
example, ``Bell's locality assumption" 
in Ref.~\cite{Aspect82} does not mean Bell-locality as per Definition~\ref{def:BN}, but rather 
(as it should, since Aspect {\em et al.} cite only Bell's 1964 paper~\cite{Bel64}), 
locality as per Definition~\ref{def:local}.}.  \azur
But, on the positive side, it is surely a term of which 
the latter-day Bell would have approved.

\section*{Acknowledgements}

I am deeply grateful for comments on drafts of this manuscript by, in chronological order 
of first communication: Eric Cavalcanti,  Eleanor Rieffel,  Michael Hall,  Matthew Pusey,  Rachael Briggs, 
 Roger Colbeck,  Travis Norsen (who wishes it to be noted that he completely disagrees with my analysis 
and finds my arguments flimsy, circular, and wholly unconvincing),  Tim Maudlin, \red and Mitchell Porter.   \blk
This research was supported by the ARC Centre of Excellence CE110001027 and 
by the ARC Discovery Project DP140100648.

\appendix

\section{EPR, Completeness, Bell's Theorem, and \red fragile \blk locality} \label{App:EPR}

In this Appendix, I summarise the formalisation in Ref.~\cite{Wis13} of the 
criteria used by EPR~\cite{EPR35}. The purpose of this material is three-fold. 
The first is to explain how, at least on my reading of EPR, Bell's 1964 theorem can be thought of 
as a proof that no theory of quantum phenomena can be EPR-complete. Thus, 
EPR's completeness is a candidate for Bell's imagined 
hypothesis H in his 1964 paper. The second is so that the reader will appreciate how subtle 
EPR-completeness is. This is to explain, in part, why it is implausible 
to imagine that Bell had this concept in mind. The third is to relate EPR-completeness 
to Jarrett-completeness, which forms the last part of the Appendix. This Appendix draws 
some of its material directly from Ref.~\cite{Wis13}.
 
\subsection{Preliminaries}

To connect the formalism of Bell's papers to the language of EPR, 
it is necessary to talk of {\em systems}. 
To give this an operational meaning (that is, one grounded in macroscopic events), 
assume that events $a$ and $A$ 
are locatable within a space time region $\alpha$ which is disjoint from another region $\beta$ containing $b$ and $B$. We can then identify Alice's and Bob's systems with 
$\alpha$ and $\beta$ respectively. While EPR, surprisingly, never mention spatial separation of their systems 
(let alone space-like separations) they do assume that their systems are ``no longer  interacting". 
For ease of comparison with Bell's work I will take this to hold because they are space-like separated. 

EPR also refer to {\em physical quantities} pertaining to systems $\alpha$ and $\beta$. I will denote these by $\hat{a} \in \mathbb{P}_\alpha$ and $\hat{b} \in \mathbb{P}_\beta$. I use a hat because in quantum mechanics physical quantities are represented by operators, but there is no implication here that quantum mechanics must be correct. Because there are many ways to measure a physical quantity, $\hat a$ is associated with an equivalence class of settings: $a \in \mathbb{S}_{\hat a}$, and similarly for $\hat b$ and $b$. 
 In what follows, unless otherwise specified, 
 the symbol $\forall\ a$ is to be understood as meaning all possible $a$; that is, $\forall\ a \in \mathbb{S}_{\hat a}$ for some $\hat a \in \mathbb{P}_\alpha$;  and likewise for $b$, $A$, and $B$, {\em mutatis mutandis}. 
 I will assume that $c$ is fixed throughout.  

\subsection{Completeness} \label{App:EPR:Com}

As with most of the concepts they introduce, EPR explain completeness reasonably precisely:
\begin{quote}
\{T\}he following requirement 
for a complete theory {seems} to be a necessary one: {\em every element of the physical reality must have a counterpart in the theory}.
\end{quote}
Here, and --- except where noted --- below, the italics are as in the original. From a careful examination of EPR's text~\cite{Wis13}, one finds that their criterion for completeness, for a theory $\theta$, should be expressed using formal logic as 
\begin{criterion}[EPR-completeness] \label{ComCriterion}
$$\Com(\theta)  \implies \sq{ \forall\  \hat b \in \mathbb{P}_\beta , \ \EPR(\hat b|c) \implies  \Rep_\theta(\hat b|c) }.$$
\end{criterion} 
Here I have used $\Rep_\theta(\hat b|c)$, standing for `is represented', to denote EPR's concept that $\hat b$ ``has a counterpart" in theory $\theta$, and $\EPR(\hat b|c)$ to mean that the property $\hat b$ is an `element of physical reality'. 

EPR do not actually define what it means for $\hat b$ to `have a counterpart in the theory', but I think it is uncontroversial to take it to imply that the outcome of any measurement of $\hat b$ is determined in the theory:
\begin{criterion}[Representation in the theory] \label{RepDef}
$$\Rep_\theta(\hat b|c) \implies \forall\ B, b, \lambda, \ P_\theta(B|b,\lambda,c) \in \cu{0,1}. $$
\end{criterion}
In other words, $\Rep_\theta(\hat b|c)$ implies that there exists a function  
$B_\theta(b,\lambda,c)$. 
Note that saying this is strictly stronger than saying that 
a measurement of $\hat b$ has predetermined outcomes, which would allow for a functional 
dependence on $a$, the measurement setting for the other system: $B_\theta(b,a,\lambda,c)$.
For an explanation as to why EPR cannot have intended $a$ to appear as an argument in their notion of 
representation in the theory, see Ref.~\cite{Wis13}. However, this explanation does not actually rule out the alternative criterion where the consequent in Criterion~\ref{RepDef} is replaced by $\forall\ a, B, b, \lambda, \ P_\theta(B|a, b,\lambda,c) \in \cu{0,1}$. Interestingly, adding the assumption of \red fragile \blk locality, as per Definition~\ref{def:fragile-locality} in Sec.~\ref{sec:JarShi}, would immediately transform this alternative criterion into Criterion~\ref{RepDef}. 

Regarding elements of physical reality, EPR give the following sufficient criterion: 
\begin{quote} 
 {\em If, without in any way disturbing a system, we can predict with certainty (\ie, with probability equal to unity) the value of a physical quantity, then there exists an element  of physical reality corresponding to this physical quantity.} 
\end{quote}
To enable a non-disturbing measurement on $\beta$ they consider an {\em indirect} measurement (\ie, a measurement on the other system $\alpha$), so formally  we have:  
\begin{criterion}[Element of Physical Reality]\label{EPRcriterion}
$$\EPR(\hat b|c) \follows \exists\  a :  [\ \neg \Dis(\beta|a,c) \et \Pre(\hat b|a,c) \ ].$$ 
\end{criterion}
Here $\Dis(\beta|a,c)$ means that the measurement $a$ disturbs the system $\beta$, while $\Pre(\hat b|a,c)$ 
means that the measurement $a$ makes the value of $\hat b$ predictable.  

Like representation, predictability can  be defined uncontroversially: 
\begin{definition}[Predictability]\label{def:Pre}
$$\Pre(\hat b|a,c) \iff \forall\  B, b,  A, \ f(B|A,a,b,c) \in \cu{0,1}.$$
\end{definition} 
Note the use here of the operationally determined relative frequencies $f$, not the theory probabilities $P_\theta$ --- 
this is what differentiates predictability from determinism; see \eg~Ref.~\cite{CavWis12}. 
The second new notion --- disturbance --- is the most problematic in EPR's paper, and 
 it was the notion of disturbance which for Bohr~\cite{Boh35a,Boh35} 
 was the key to refuting EPR, as he 
rejected their notion and proposed his own, clearly stated, sufficient condition~\cite{Wis13}. 
Regardless of how EPR understood disturbance intuitively, it is apparent from the form of 
their argument that they thought it obvious that a measurement on system $\alpha$ would not disturb 
system $\beta$ if the systems were no longer interacting.  Since throughout we are taking $\alpha$ and $\beta$ to be space-like separated, no disturbance is  (for EPR but not for Bohr) a property of all indirect measurements: 
 \begin{assumption}[No Disturbance] \label{defDis1}
$$\forall a ,\ \neg \Dis^{\rm EPR}(\beta|a,c)$$
\end{assumption}

\subsection{From Completeness to Bell's Theorem} \label{App:EPR:Bel}

If we accept the criteria given by EPR as set out above, 
and EPR's assumption \ref{defDis1}, 
we can follow EPRÕs argument to conclude that
  \beq \label{EPRgoal}
 \Com(\theta) \et [\ \exists\  a :  \Pre(\hat b|a,c) \ ] \implies \Rep(\hat b|c).
 \eeq
 As noted above, $\Rep(\hat b|c)$ is equivalent to the local predetermination 
 of all outcomes $B$, 
 which is what Bell claimed to have  derived  in 1964 from the assumption of locality. 
 Here we see that $\Com(\theta)$ works, where locality does not. 
 Clearly EPR-completeness involves localistic notions, in the functional dependencies of 
 $\Rep(\hat b|c)$ and $\EPR(\hat b|c)$ and in the assumption that $\neg \Dis(\beta|a,c)$, 
 but EPR never explicitly appeal to locality as defined by Bell in 1964, 
 or to local causality as defined by Bell in 1976. At the end of the paper, EPR 
 entertain, but then reject, the possibility of allowing $\EPR(\hat b|a,c)$: 
 \begin{quote}
 \{T\}his point of view \ldots\ makes the reality of $P$ and $Q$ 
 \{(properties $\hat b$ of the first system)\}  depend upon the process of measurement 
 \{($a$)\} carried out on the first system, which does not disturb the second system in any way. 
 No reasonable definition of reality could be expected to permit this. 
 \end{quote}
 While this quote certainly suggests that EPR would consider locality as per Definition~\ref{def:local} 
 to be a {\em sine qua non}, it is not an explicit assumption which they use in conjunction with 
 other assumptions to derive their result. Rather, it is built into the very structure of their 
 criteria~\cite{Wis13}. 
 
  If we say that, in assuming EPR's notion of completeness, we 
 accept all of EPR's criteria and assumptions, then we can, following their argument, 
 prove their theorem: 
 \begin{theorem}[EPR] \label{thm:EPR1}
There exist quantum phenomena for which OQM is not an EPR-complete theory.  
 \end{theorem} 
 Moreover, assuming completeness in this way,  for the singlet state and projective measurements 
 we can follow Bell's argument to 
 derive that any spin observable, for either party, must be an $\EPR$, and hence that it must have a counterpart in the theory, that is, be locally predetermined. But Bell's theorem says that any theory satisfying 
 L\&D fails to predict certain correlations on the singlet state.  Thus we could state 
 \begin{theorem}[Bell 1964 as EPR-completeness] \label{thm:Bell1}
There exist quantum phenomena for which there is no EPR-complete theory.  
 \end{theorem} 
 
 Note that this statement of Bell's theorem can only be proven for a joint system with perfect correlations, 
 as in a spin-singlet state with projective measurements. Without perfect predictability, EPR's network of 
conditions literally lead nowhere. But perfect correlation is an idealisation that is never seen in experiment. 
 Hence, while Theorem~\ref{thm:Bell1} is formally valid, one cannot conclude from it that 
there exist {\em phenomena in the world} for which there is no EPR-complete theory.  The same criticism applies to Bell's original proof of theorem \ref{thm:Bell64} in 1964. However the assumptions Bell made for this theorem, 
L\&D rather than completeness, enables other proofs that are not subject to this criticism. 
That is, the assumption of L\&D does give testable predictions even when the correlation is not perfect, 
as proven later by Clauser \ea~\cite{CHSH69}.

 \subsection{EPR-Completeness versus Jarrett-Completeness} \label{App:EPR:Jar}

In this section I prove Theorem~\ref{thm:JCFLEC}, that whatever is implied by assuming 
$\Com(\theta)$ is also implied by assuming that $\theta$ respects Jarrett-completeness and 
\red fragile \blk locality. 

First recall that, as in \erf{EPRgoal}, there 
is only one implication of assuming $\Com(\theta)$, namely 
 \beq \label{EPRgoal1}
 \forall\  \hat b \in \mathbb{P}_\beta , \ [\ \exists\  a :  \Pre(\hat b|a,c) \ ] 
 \implies  \Rep_\theta(\hat b|c) 
 \eeq
 Using Criteria~\ref{def:Pre} and \ref{RepDef}, this means
 \begin{eqnarray} \label{EPRgoal2} 
 \forall\  \hat b \in \mathbb{P}_\beta , \ [\ \exists\  a : &&\forall\  A, B, b, \ f(B|A,a,b,c) \in \cu{0,1} \ ] \nn \\
 && \implies  [\ \forall\ B, b, \lambda, \ P_\theta(B|b,\lambda,c) \in \cu{0,1} \ ].
 \end{eqnarray}
I now show the same implication follows from JC\&{\red FL},  
being more careful with quantifiers than in the main text. 

Conjoining Jarrett-completeness, 
$$
	\forall\  A, a, B, b, \lambda, \ P_\theta(B|A, a, b, c, \lambda) = P_\theta(B|a, b, c, \lambda),
	$$
and \red fragile \blk locality 
$$ 
\forall\  a, B, b, \lambda, \ [\ P_\theta(B|a, b, c, \lambda) \in \{0,1\} \implies P_\theta(B|a, b, c, \lambda) = P_\theta(B|b, c, \lambda) \ ],
$$
we have
$$
\forall\  A, a, B, b, \lambda, \ [\ P_\theta(B|A, a, b, c, \lambda)\in \{0,1\}  \implies P_\theta(B|b, c, \lambda) \in \{0,1\} \ ]. 
$$
Now it is true by definition that  
$$\forall\  A, a, B, b, \lambda, \ [\ f(B|A, a, b, c) \in \{0,1\} \implies P_\theta(B|A, a, b, c, \lambda)\in \{0,1\} \ ],$$
so we have
\beq
\forall\  A, a, B, b, \lambda, \ [\ f(B|A, a, b, c)\in \{0,1\}  \implies P_\theta(B|b, c, \lambda) \in \{0,1\} \ ]. \label{JCFL-strong}
\eeq
We can weaken \erf{JCFL-strong} to 
$$
\forall\  \hat b \in \mathbb{P}_\beta ,  \ \forall\  a, \ [ \ [\ \forall\ A, B, b,\ f(B|A, a, b, c)\in \{0,1\}]   \implies [\ 
\forall\ B, b,\ P_\theta(B|b, c, \lambda) \in \{0,1\} \ ] \ ],
$$
which is the same as \erf{EPRgoal2}.  In other words, 
the only implication that from EPR's network of criteria, \erf{EPRgoal2}, 
can be derived from the assumption that $\theta$ \red satisfies \blk JC and {\red FL}. 
This completes the proof. 


\section*{References}
\input{2Bells_2014n-arxiv.bbl}


\end{document}

%% file: 2Bells_2014n-arxiv.bbl
\providecommand{\newblock}{}